\documentclass[fleqn,usenatbib]{mnras}

\usepackage{newtxtext,newtxmath}
\usepackage{anyfontsize}
\usepackage{soul}

\usepackage[T1]{fontenc}

\DeclareRobustCommand{\VAN}[3]{#2}
\let\VANthebibliography\thebibliography
\def\thebibliography{\DeclareRobustCommand{\VAN}[3]{##3}\VANthebibliography}

\usepackage{graphicx}	
\usepackage{amsmath}	
\usepackage{siunitx}
\usepackage{orcidlink}
\usepackage{longtable}

\title[Photometric Variability Towards M3 with TWIST]{A search for photometric variability towards the globular cluster M3 with the TWenty Inch Survey Telescope}

\author[M. A. Mitchell et al.]{Morgan A. Mitchell$^{\orcidlink{0009-0004-6130-7775}}$,$^{1,2}$\thanks{E-mail: morgan.mitchell@warwick.ac.uk}
    Paul Chote,$^{1,3}$
    James McCormac$^{\orcidlink{0000-0003-1631-4170}}$,$^{1,2,3}$
    Don Pollacco$^{\orcidlink{0000-0001-9850-9697}}$,$^{1,2,3}$
    \newauthor
    Ioannis Apergis$^{\orcidlink{0009-0004-7473-4573}}$,$^{1,2}$
    J. D. Lyman$^{\orcidlink{0000-0002-3464-0642}}$,$^{1}$
    Isobel S. Lockley$^{\orcidlink{0009-0003-0928-3588}}$,$^{1,2}$
    Samuel Gill$^{\orcidlink{0000-0002-4259-0155}}$,$^{1,2}$
    James A. Blake$^{\orcidlink{0000-0002-5903-2387}}$,$^{1,3}$
    \newauthor
    Alastair B. Claringbold$^{\orcidlink{0000-0003-1309-5558}}$,$^{1,2}$
    D. T. H. Steeghs$^{\orcidlink{0000-0003-0771-4746}}$,$^1$ and J. Casares$^{\orcidlink{0000-0001-5031-0128}}$ $^{4,5}$
\\
$^{1}$Department of Physics, University of Warwick, Gibbet Hill Road, Coventry CV4 7AL, UK \\
$^{2}$Centre for Exoplanets and Habitability, University of Warwick, Gibbet Hill Road, Coventry CV4 7AL, UK \\
$^{3}$Centre for Space Domain Awareness, University of Warwick, Gibbet Hill Road, Coventry CV4 7AL, UK \\
$^{4}$Instituto de Astrofísica de Canarias, E-38205 La Laguna, Tenerife, Spain \\
$^{5}$Departamento de Astrofísica, Universidad de La Laguna, E-38206 La Laguna, Tenerife, Spain \\
}

\date{Accepted 2025 November 24. Received 2025 November 17; in original form 2025 September 12}

\pubyear{2025}

\begin{document}
\label{firstpage}
\pagerange{\pageref{firstpage}--\pageref{lastpage}}
\maketitle

\begin{abstract}

We present the commissioning results and first scientific observations from the newly installed TWIST observatory -- a 50\,cm telescope equipped with an sCMOS camera providing a $36.1\times24.1$\,arcmin$^2$ field of view -- housed in the former SuperWASP-North enclosure. We conducted a 67-night, 199-day baseline white-light monitoring campaign centred on the globular cluster M3 aimed at characterizing stellar variability within the cluster while also assessing the photometric performance of the newly commissioned system. We report the discovery of four new SX Phoenicis variables (V301--304), confirm their cluster membership, and identify fundamental-mode pulsation in one, allowing an independent period-luminosity-based distance estimate to M3. We revisited 231 previously known RR Lyrae stars, providing updated period measurements for 203 and white-light amplitudes for 198. We detected Blazhko-like modulation in 53 stars and characterized the modulation parameters for 28. Notably, we measure periods and amplitudes for the unclassified variables V286 and V287 for the first time. We also identify three foreground flaring M dwarfs, and assess the feasibility of detecting microlensing events in M3, concluding that expected rates are negligible. Alongside the scientific results, we introduce a new correction technique for flat-field images affected by scattered light and present a full characterization of the observatory's photometric capabilities. These results demonstrate the scientific utility of TWIST for high-cadence time-domain surveys using modest-aperture instrumentation.
\end{abstract}




\section{Introduction}

The globular cluster M3 (NGC 5272) is among the most extensively studied in the Milky Way. Located at a distance of 10.2 kpc \citep{harris_catalog_1996} at high Galactic latitude ($b = 78.7^\circ$), M3 contains one of the richest populations of variable stars in any Galactic cluster, second only to $\omega$~Cen. Since the discovery of the first periodic variable in a globular cluster by \citet{pickering-variable}, M3 has served as a benchmark for studies of RR Lyrae variables and horizontal branch evolution \citep[e.g.][]{catelan_evolutionary_2004}. The most recent update to the catalogue by \citet{clement_variable_2001} in March 2019 lists 273 confirmed variables, including 241 RR Lyrae (RRL) and 9 SX Phoenicis (SXP) stars.

M3 is the prototypical Oosterhoff type I cluster \citep{oosterhoff_remarks_1939}, exhibiting relatively short-period RR Lyrae variables and a greater predominance of fundamental-mode (RR\textit{ab}) pulsators. While once considered a classic example of a simple stellar population, M3 is now known to host multiple populations \citep[e.g.][]{massari_multiple_2016, lee_multiple_2021}, possibly the result of a merger. The cluster also hosts the first discovered blue straggler stars \citep{sandage_color-magnitude_1953}, of which its SXP variables are examples. The radial distribution of M3's blue stragglers shows a bimodal structure \citep[e.g.][]{ferraro_blue_1993, ferraro_hst_1997}, and while early surveys found relatively few SXPs \citep{kaluzny_search_1998, hartman_bvi_2005}, this may reflect selection biases in observing cadence and sensitivity.

M3 has remained the focus of ground-based variability studies for the last several decades across a wavelength range spanning the near-ultraviolet to the near-infrared. Early modern near-infrared observations by \citet{longmore_globular_1990} explored the \textit{K}-band period--luminosity relation of RR Lyrae variables in M3 and other clusters, demonstrating a correlation between $\log{(\mathrm{period})}$ and \textit{K} magnitude almost free of systematic scatter. Subsequent CCD variability surveys were undertaken by \citet{kaluzny_search_1998} in the \textit{V}-band, \citet{corwin_bv_2001} in the \textit{BV}-bands and \citet{hartman_bvi_2005, benko_multicolour_2006} in the \textit{BVI}-bands. They obtained high-quality light curves for hundreds of variables, including newly identified RRL and SXP stars. These works presented 9, 14, 19 and 9 nights of data, respectively, and had minimum observing cadences ranging from around 5 to 20 minutes.

A series of follow-up investigations by \citet{jurcsik_long-term_2012, jurcsik_overtone_2015, jurcsik_photometric_2017, jurcsik_blazhko-type_2019} further examined the long-term stability, mode switching, Blazhko modulation, and physical properties of M3’s RR Lyrae population through extended photometric and spectroscopic monitoring. In the near-infrared, \citet{bhardwaj_near-infrared_2020} obtained time-series $JHK_s$ photometry for 233 RR Lyrae stars, establishing precise mean magnitudes and updated period--luminosity relations. Most recently, \citet{kumar_multiwavelength_2024} presented a multiwavelength ($UBVRIJHK_s$) analysis based on an extensive 35-year dataset, deriving improved periods, Fourier-based light-curve parameters, and inferring corresponding physical properties using an artificial neural network for a large sample of RR Lyrae stars.

The aforementioned variability studies have all used CCD imaging with cadences of minutes, which may limit sensitivity to short-period pulsators and brief transients. Recent advances in scientific complementary metal-oxide-semiconductor (sCMOS) technology now offer an alternative to CCDs for time-series photometry. With improved quantum efficiency, lower read noise, and orders of magnitude shorter readout times, sCMOS cameras are increasingly adopted in both ground-based projects \citep[e.g.,][]{licandro_atlas-teide_2022, law_low-cost_2022, AIREY20255757} and proposed space missions \citep[e.g.,][]{malbet_theia_2022, ge_et_2022}. Large-format sCMOS sensors are now readily available at a lower cost than many traditional CCDs. These capabilities make sCMOS cameras particularly well-suited for high-cadence, wide-field photometry -- ideal for monitoring dense stellar fields such as globular clusters. However, their performance in long-term, time-domain studies remains relatively underexplored.

In this paper, we present the first results from TWIST (the TWenty Inch Survey Telescope), a new 50 cm telescope equipped with an sCMOS camera, housed in the former enclosure of the SuperWASP-North facility. We used TWIST to carry out a 10-second cadence, 67-night, 199-day baseline white-light photometric campaign of M3, aimed at validating the system's photometric capabilities and evaluating its scientific return. 

We introduce TWIST in Section \ref{sec:twist}, and describe the observations and data reduction, including a new method for correcting scattered-light contamination in flat-field images in Section \ref{sec:obs}. We report new discoveries and updated properties for known variable stars in M3 in Section \ref{sec:freq_analysis}, and evaluate the probability of observing microlensing events originating from the cluster in Section \ref{sec:non-periodic}. In Appendices \ref{app:characterization} and \ref{app:photometric_precision}, we assess the performance of the observatory.

\section{TWIST}
\label{sec:twist}

As this study represents the first scientific use of the TWenty-Inch Survey Telescope (TWIST), we briefly describe its setup and properties here. Detailed characterization of the instrument and photometric performance is provided in the Appendices.

TWIST (shown in Fig.~\ref{fig:TWIST}) is a general-purpose, fully robotic imaging telescope equipped with a full-frame, backside-illuminated sCMOS detector. Operated and maintained by the University of Warwick, it has been in operation since its installation in March 2023 at the Observatorio del Roque de los Muchachos on La Palma, Canary Islands. The telescope is housed in the former enclosure of the SuperWASP-N observatory, which uses a sliding roof design. The building is outfitted with a Vaisala\footnote{https://www.vaisala.com/en} weather station and multiple rain detectors, as well as an uninterruptible power supply (UPS) to ensure the roof can close in the case of loss of power. Additional details about the enclosure and its design can be found in \cite{pollacco_wasp_2006}.

\begin{figure}
    \begin{center}
    \includegraphics[width=85mm]{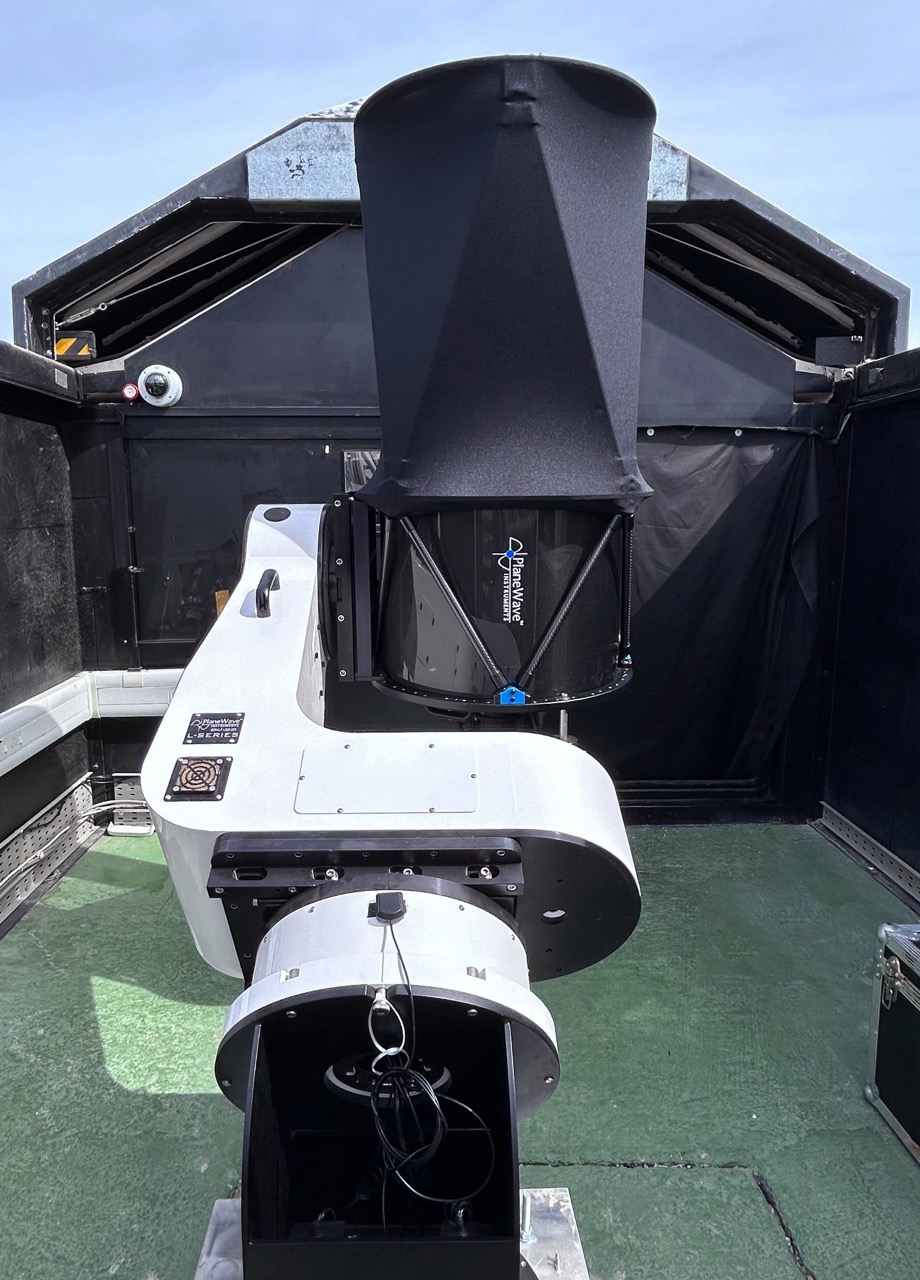}
    \end{center}
    \caption{The TWenty-Inch Survey Telescope in its enclosure on La Palma, Canary Islands. This image was taken from the southern entrance of the building, facing North.}
    \label{fig:TWIST}
\end{figure}

\subsection{Telescope}
\label{subsec:telescope}

TWIST comprises a commercially available PlaneWave Instruments\footnote{https://planewave.com/} CDK20 optical tube assembly (OTA) mounted equatorially on a PlaneWave Instruments L500 direct drive mount, which does not require a pier flip when crossing the meridian. The OTA is of the "Corrected Dall-Kirkham" design, which uses an ellipsoidal primary mirror, a spherical secondary mirror, and two corrector lenses. PlaneWave Instruments claims this produces a flat focal plane and minimal off-axis coma and astigmatism up to a 52\,mm imaging circle. This OTA uses an open truss design that permits light to be scattered onto the optics through the gaps of the trusses, so a light shroud was installed in February 2024 to prevent stray light from entering the optical path. PlaneWave Instruments describes the CDK20 as a "diffraction-limited telescope". The complete hardware specifications of the system are detailed in Table \ref{tab:specs}.

\begin{table}
	\centering
	\caption{Details and specifications of TWIST's optical tube assembly, mount and camera. The camera read noise, full well capacity, dynamic range, gain, and linearity error were all measured in the camera's "Photographic" mode at gain setting 26 -- the setup used for the survey in the present work. The readout time is applicable to the streaming frame transfer mode.}
	\label{tab:specs}
	\begin{tabular*}{\linewidth}{@{\extracolsep{\fill}} ll }
		\hline
		  \textbf{Quantity} & \textbf{Value} \\
		\hline
            \textbf{Telescope}\\
            Latitude & 28$^{\mathrm{\circ}}$45\arcmin36\farcs9 N\\
            Longitude & 17$^{\mathrm{\circ}}$52\arcmin45\farcs3 W\\
            Altitude & 2349\,m \\
            Primary diameter & 508\,mm \\
            Secondary diameter & 191\,mm \\
            Focal length & 3454\,mm \\
            Focal ratio & $f$/6.8 \\
            Plate scale & 59.7\,arcsec\,mm$^{-1}$ \\
            Slew speed & 50\,deg\,s$^{-1}$ \\
            \hline
            \textbf{Camera} \\
            Total pixel area & 9600\,$\times$\,6422 \\
            Exposed pixel area & 9576\,$\times$\,6388 \\ 
            Pixel size & 3.76\,$\unit{\um}$ \\
            Pixel angular scale & 0.225\,arcsec\,pixel$^{-1}$ \\
            Camera field of view & 36.1\,$\times$\,24.1 arcmin$^2$ \\
            Read noise (RMS) & 2.579\,e$^-$ \\
            Full well capacity & 22200\,e$^-$ \\
            Dynamic range & 78.7\,dB \\
            Gain & 0.414\,e$^-$ADU$^{-1}$ \\
            Linearity error (EMVA 1288 v4.0*) & 0.248\,$\%$ \\
            Dark current (0$^\circ$C) & 8.25\,e$^-$pixel$^{-1}$hour$^{-1}$ \\
            Readout time & 39.027\,$\unit{\us}$ \\
		  \hline
          \multicolumn{2}{l}{\small *https://www.emva.org/standards-technology/emva-1288/} \\

	\end{tabular*}
\end{table}

\subsection{Camera}
\label{subsec:camera}

TWIST is equipped with a QHYCCD\footnote{https://www.qhyccd.com/} QHY600M Pro sCMOS camera, a motorized electronic focuser, and a five-slot filter wheel housing a blocking filter for taking bias and dark images, as well as Bessel \textit{B}, \textit{V}, and \textit{R} filters; one slot remains open for white-light photometry. Combined with the optical tube, the $36\times24$\,mm$^2$ Sony IMX455 sensor of the QHY600M Pro produces an on-sky field of view of $36.1\times24.1$\,arcmin$^2$. A detailed characterization of the camera's performance and the total system spectral response is presented in Appendix~\ref{app:characterization}, and the key results from these analyses are summarized in the lower half of Table~\ref{tab:specs}. Additionally, an analysis of the photometric performance of the system is presented in Appendix~\ref{app:photometric_precision}. The photometric performance is in line with expectations for the typical observed point spread function half-flux diameter of around 3\,arcsec.

\subsection{Software}
\label{subsec:software}

The mount, camera, filter wheel, focuser, and roof are all controlled by an on-site computer running custom \textsc{rockit}\footnote{https://rockit-astro.github.io/} software on Rocky Linux\footnote{https://rockylinux.org/}. The \textsc{rockit} system is a modular observatory control framework designed to manage all aspects of robotic telescope operations, including hardware control, environmental monitoring, and data acquisition. It operates as a collection of lightweight services (daemons) that communicate through a remote procedure call interface, ensuring scalability and modularity. These services handle tasks such as aggregating weather data, managing dome and telescope hardware, automating observation schedules, and processing image metadata. In addition, a centralized web dashboard provides real-time system monitoring and access to weather, operational status, and video feeds.

This system automates flat-field imaging (morning and evening), focusing, target field acquisition, filter selection, and target observation. These operations are defined as \textit{actions} within \textsc{rockit} and can be scheduled nightly using an observing script file. The schedule is carried out for as long as the environment is considered safe. That is, the temperature, humidity, wind speed, and sun altitude are within an acceptable range, rain is not detected, the UPS is powered, network connectivity is maintained, and disk space is available. If these conditions are violated, the roof is automatically closed, and observations are suspended. These provisions negate the need for human monitoring of the observatory during its operation.

We perform autoguiding on the science images using \textsc{donuts} \citep{mccormac_donuts_2013}, allowing us to forgo a separate guide telescope, which would add unnecessary complexity.

\subsection{Visibility}
\label{subsec:visibility}

\begin{figure*}
        \includegraphics[width=9cm]{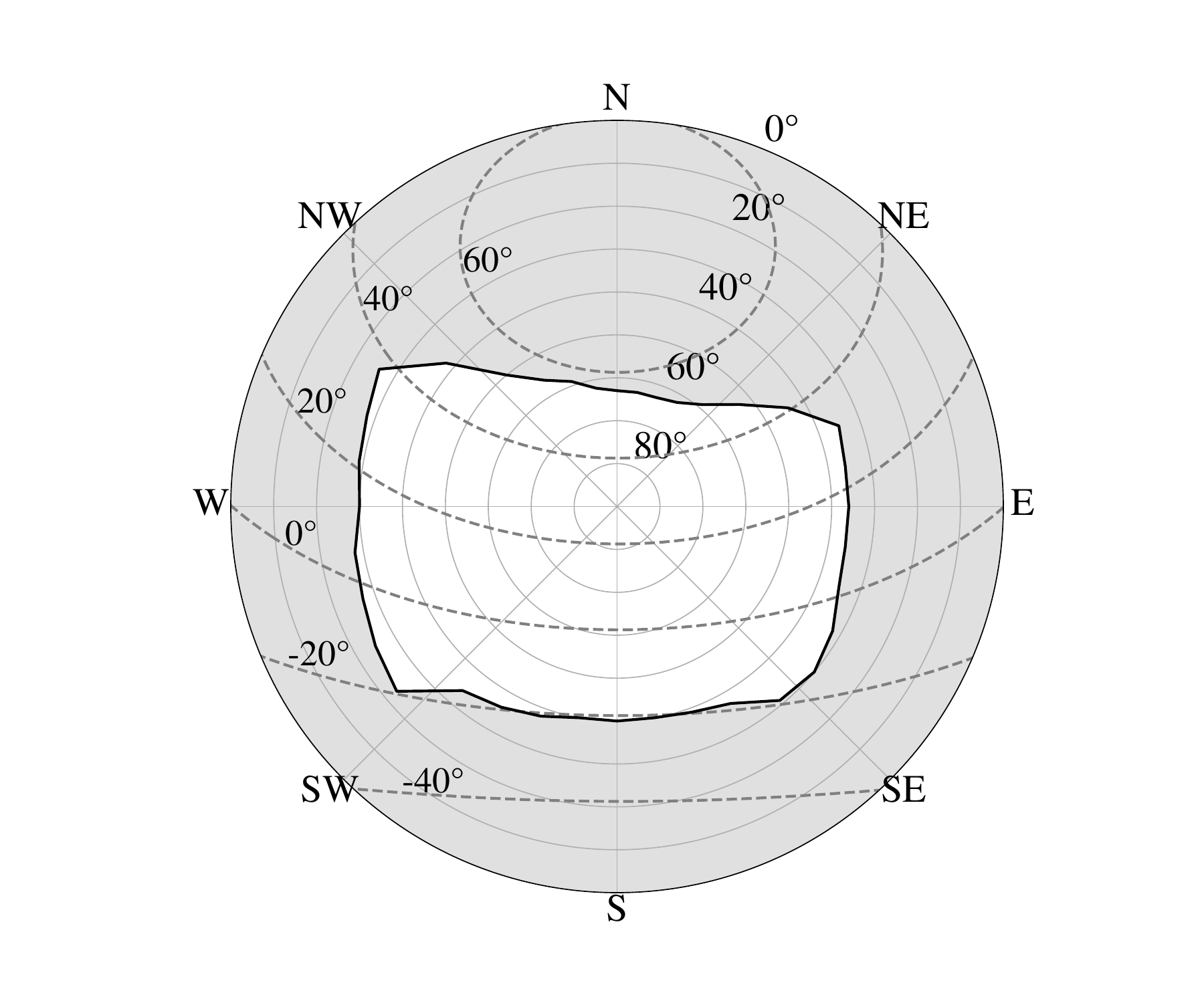}
        \includegraphics[width=8.5cm]{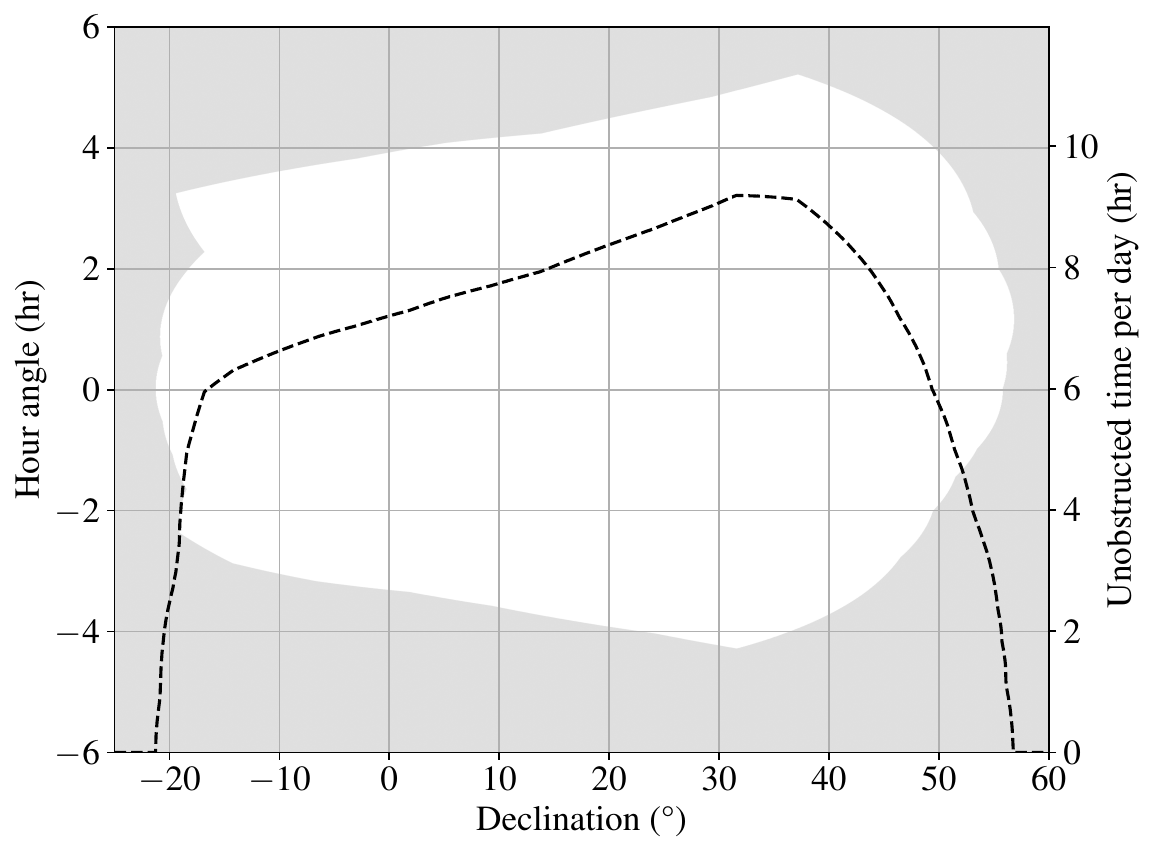}
    \caption{The observing limits of TWIST. \textbf{Left:} The minimum observable altitude (black line) measured to the nearest degree versus the azimuthal angle, sampled every 10$^\circ$. The white-shaded region shows the observable altitude/azimuth combinations. The concentric circles are lines of constant altitude and the polar angle represents the azimuthal angle. The dashed grey lines are loci of points of a fixed declination, illustrating the sidereal motion of an object of that declination. \textbf{Right:} The observable hour angles (shaded white) visible for a given declination and the total unobstructed time per night (black dashed line) at that declination. This time does not include sun or moon-induced visibility restrictions.}
    \label{fig:observability}
\end{figure*}

The sliding roof design was initially chosen to facilitate rapid movements of the SuperWASP cameras. However, this configuration limits TWIST's visibility, especially in the North, where altitudes <63$^\circ$ are obstructed by the building. Nevertheless, all objects with declinations $-22^\circ\le\delta\le56^\circ$ retain some visibility from TWIST. Additionally, the visibility is asymmetric in the East-West direction due to the L-mount design (see Fig.~\ref{fig:TWIST}). The telescope hangs lower when pointing East, and is therefore more obstructed by the walls of the enclosure than when pointing West. As a result, the positive hour angle limit is greater than the negative hour angle limit in magnitude for a given declination. The full observing constraints of TWIST are illustrated in Fig.~\ref{fig:observability}. 

\section{Observations and Data Reduction}
\label{sec:obs}

Once TWIST became operational, we undertook a dedicated survey of the large, variable-rich galactic globular cluster M3 as its first science target.

\subsection{Observations}
\label{subsec:obs}

\begin{figure}
    \begin{center}
    \includegraphics[width=85mm]{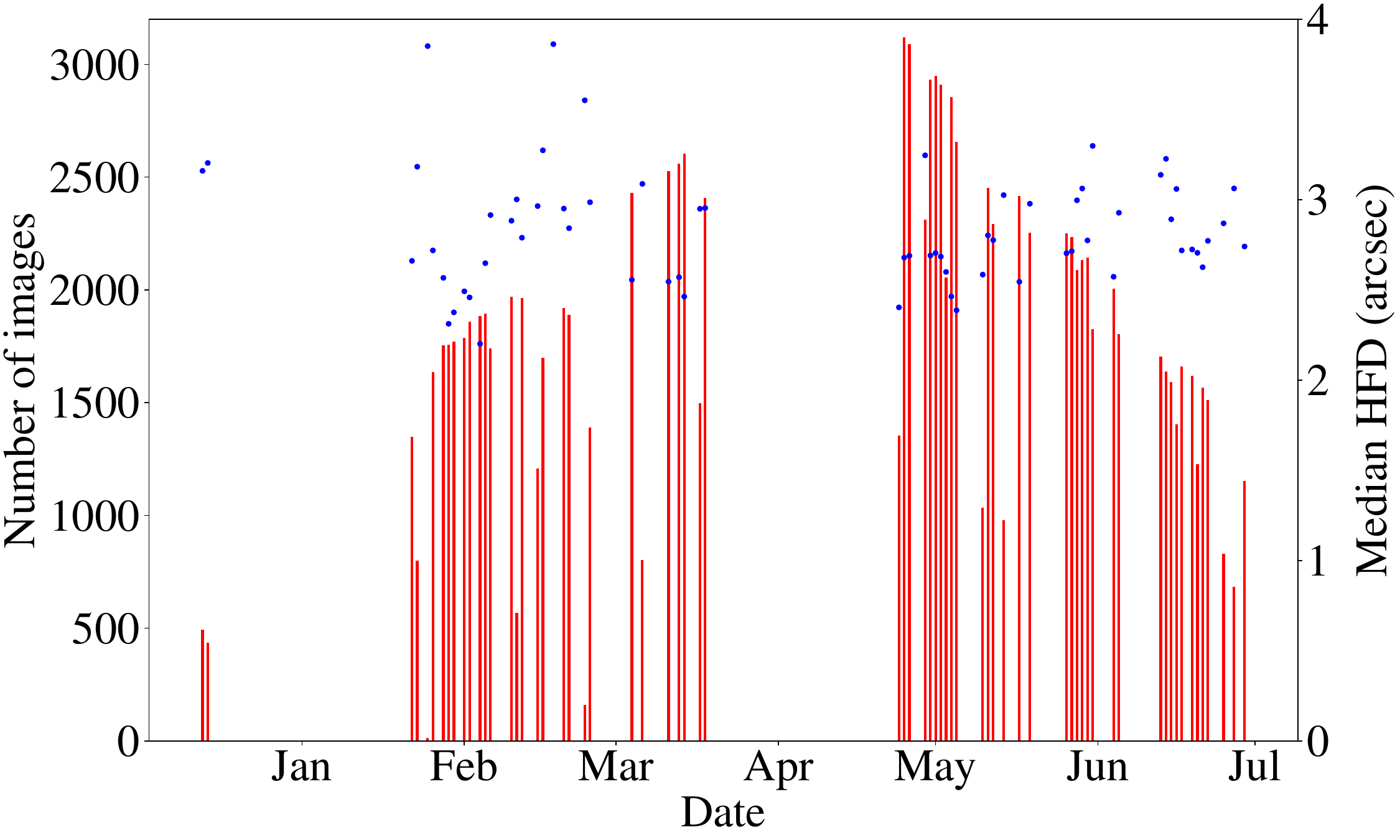}
    \end{center}
    \caption{Overview of the M3 survey observations. The red bars represent the total number of 10-second exposure, on-target images captured each night, and the blue points indicate the median half-flux diameters of all images taken on that night. Nights with a lower-than-expected number of images typically reflect either delayed starts to observations or early terminations due to environmental triggers. The two large gaps in the data are the result of deliberate non-observing periods, not adverse weather or other external factors. The month labels reflect the beginning of the month in the year 2024.}
    \label{fig:obs}
\end{figure}

We conducted initial test observations of M3 on 2023 December 13–14, when nightly visibility was approximately one hour. Regular monitoring commenced on 2024 January 22, with observations carried out on every clear night. We produced custom code to calculate the observable window each night, folding in the complex observing constraints of TWIST. This code also automatically generated the nightly observing script. When environmental constraints allowed, we observed each night for the entire visible window and took morning and evening flats. Operations were paused between 2024 March 18 and 2024 April 24 for system maintenance, and the survey culminated on 2024 June 29 when nightly visibility fell below three hours. The survey observations are overviewed in Fig. \ref{fig:obs}. In total, 130,074 science images were captured, corresponding to over 15 days of exposure time.

\begin{figure*}
    \begin{center}
    \includegraphics[width=170mm]{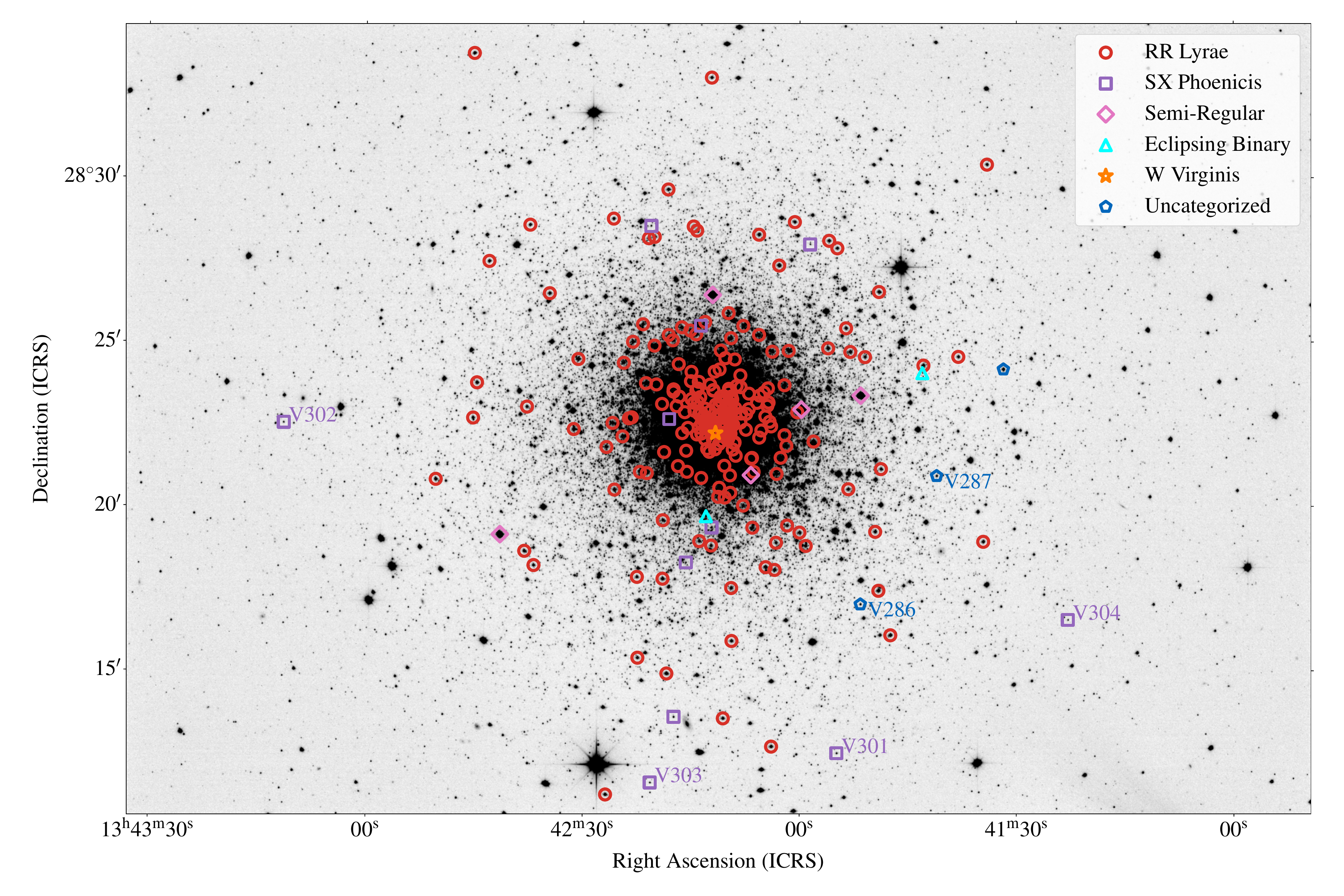}
    \end{center}
    \caption{The $36.1\times24.1$\,arcmin$^2$ survey field centred on M3. The locations of known variables from \protect\cite{clement_variable_2001} that have a corresponding \textit{Gaia} source are shown. RRL, SXP, semi-regular, eclipsing binary, W Virginis and uncategorized stars are shown with open circles, squares, diamonds, triangles, stars, and pentagons, respectively. The newly discovered SXP-type variables V301--304 and the unclassified variables V286-287 are labelled. North is up and East to the left. The image was created by aligning with \textsc{spalipy} \protect\cite[]{lyman_spalipy_2021} and stacking 360 calibrated frames (totalling one hour of exposure) with measured half-flux diameters less than 2\,arcsec taken during dark sky environments. The 5$\sigma$ limiting magnitude of the stacked image for a 5-pixel aperture radius is around \textit{G}\,=\,22.1.}
    \label{fig:field}
\end{figure*}

M3 has a tidal radius of 28.7\,arcmin \citep{harris_catalog_1996}, comparable to the $36.1\times24.1$\,arcmin$^2$ field of view of TWIST. Thus, the overwhelming majority of the cluster’s stars are encompassed within a single pointing. The field was centred on the cluster’s core to maximize the number of member stars included. The full field is shown in Fig.~\ref{fig:field}. 

To avoid excessive data rates, we binned the pixels to $2 \times 2$ bins, creating an effective angular pixel scale of 0.45\,arcsec\,pixel$^{-1}$, which still oversamples the typical seeing. To ensure the efficacy of the autoguiding algorithm, we limited exposure times to 10\,s in this work. We opted to forgo a photometric filter in order to maximize the amount of light collected -- given TWIST's relatively small aperture primary mirror -- in order to observe as deep as possible. We ran the camera in its streaming frame transfer mode, where the readout time was only 39.027\,$\unit{\us}$, corresponding to a fraction of time spent actively exposing (duty cycle) of over 99.9996 per cent. Finally, we operated the camera in its "Photographic" mode with gain setting 26 -- a configuration that moderately favours lower read noise at the expense of reduced full-well capacity. This compromise was chosen to balance noise performance and dynamic range, given the faintness of our targets -- the apparent \textit{Gaia} DR3 \citep{gaiadr3} \textit{G}-band magnitude of the horizontal branch is around 15.5\,mag, while the main-sequence turn-off occurs at around 19\,mag.

When conditions allowed, around 80 flat-field images were taken under the automated control of the \textsc{rockit} system per morning/evening twilight. Images were taken at 75$^\circ$ altitude and anti-solar azimuth at sun altitudes less than $-4^\circ$. The exposure times were automatically adjusted to maintain $\sim$50 per cent digital saturation up to a maximum exposure time of 10\,s. At the end of each night, the camera was automatically shut down, and the telescope was parked facing West at an altitude of 20$^\circ$ to both prevent debris from falling on the telescope optics and to ensure the telescope cannot point at the sun in case the roof fails to close.

\subsection{Data Reduction}
\label{subsec:reduction}

\subsubsection{Master Dark and Bias Frames and the Bad Pixel Map}

With the same camera settings used for the survey and the blocking filter in place, we took 21 dark images, each with a 300\,s exposure, and 21 bias images with the minimum 39\,$\unit{\us}$ exposure. We mean-averaged the bias images to create the master bias frame. We mean-averaged the dark frames and subtracted the master bias to create the master dark frame used in the subsequent data reduction. From the master dark image, we created a hot pixel map of pixels with greater than 1000 ADU, roughly corresponding to the transition from read-noise-dominated to dark-current-dominated noise. This method identified 40986 pixels (0.27 per cent of the total) as hot according to this definition. In addition, we used \textsc{ccdproc}'s \citep{craig_ccdproc_2015} \textsc{ccdmask} function on an evening of flat-field images to identify pixels with a non-linear photo response. This identified 2054 pixels, of which 1011 were already present in the hot pixel map. The union of these maps resulted in the combined bad-pixel map.

\subsubsection{Flat Field Corrections}

To account for the regular appearance and disappearance of dust grains on the telescope's optical components (most visibly on the camera window), we created a new master flat frame for every night on which valid (greater than 30$^\circ$ lunar separation) flat field images were taken. The master flat frames were constructed by median stacking the individual dark- and bias-subtracted flat images which were themselves normalized by their median pixel values. With this method, we noted significant variations in the vignetting pattern of the master flat from night to night and especially from dusk to dawn, which we attributed to scattered light entering the optical system. In our original analysis, these variations manifested as jumps in the photometry from one night to the next at the $\sim$1 per cent level.

\begin{figure*}
	\includegraphics[width=17.5cm]{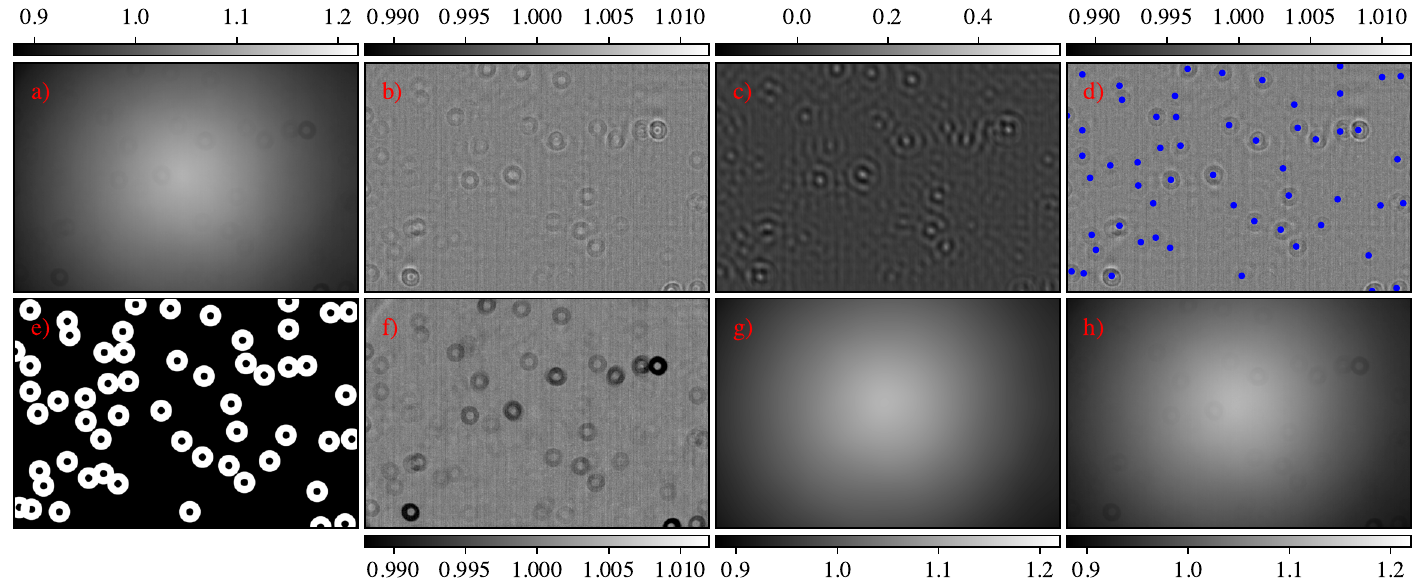}
    \caption{Overview of the flat-field correcting process. \textbf{a)} The untouched master flat from 2023 December 13 -- the first night of observations. \textbf{b)} The flat image detrended by a bivariate spline fit. \textbf{c)} Cross-correlation map of the spline-detrended flat with an example dust doughnut. \textbf{d)} The spline-detrended flat overlayed with the locations of the detected doughnuts. \textbf{e)} The fit mask; masked areas are shown in white. \textbf{f)} The RBF-detrended master flat. \textbf{g)} The master vignetting pattern used for the whole survey. \textbf{h)} The corrected master flat image.}
    \label{fig:flat_detrend}
\end{figure*}

While we expect the pattern of dust and debris in the flat images to evolve with time, the overall vignetting pattern is simply a product of the fixed optical geometry of the system -- barring minute changes arising from focus shifts or temperature changes -- and thus should remain constant. Thus, we devised a method whereby we fit and detrend the vignetting pattern of the master flats, while leaving the dust spots and the photo response non-uniformity (PRNU) information contained in the flats. This way, we can apply the same vignetting pattern to every master flat, preventing night-to-night variations while being sensitive to the appearance and disappearance of dust spots and a potentially evolving PRNU. The correcting process is illustrated in Fig. \ref{fig:flat_detrend}.

Dust and debris on the camera window create doughnut shapes in the flat field images with lower counts than the surrounding area. Masking these doughnuts from the fit was essential to capturing the underlying vignetting trend. For each night, we first detrended the master flat with a rectangular bivariate spline using \textsc{scipy} \citep{2020SciPy}. We then cross-correlated the resulting image with a sharp, uncontaminated example of a dust doughnut to produce a cross-correlation map. The locations of the dust grains were then inferred from the positions of the peaks on the map. This approach leverages the fact that all the doughnuts are similar in size and shape. The dust locations were used to generate a fit mask, conservatively approximating the doughnuts as the region between two concentric circles with radii of 50 and 150 pixels, respectively. We then used the \textsc{rbf} package of \cite{rbf} to interpolate the masked master flat with radial basis functions (RBFs). The final trend was calculated as the result of mean-averaging 30 interpolants, each generated using 1,000 samples of the unmasked region of the image.

After each night's master flat was detrended, we restored the vignetting by multiplying by the trend from the flats of 2024 April 24 -- the first night after which the camera window had been cleaned during the survey, and by which time the stray light-blocking truss sleeve had been installed -- we refer to this as the master vignetting pattern. For uniformity, this master trend was used for all nights in the dataset. Where no valid flats were taken, the master flat frame from the closest available night was used.

The value of the RBF approach to detrending over a simple spline fit is demonstrated in panels \textbf{b)} and \textbf{f)} of Fig. \ref{fig:flat_detrend} by the lack of artefacts around the doughnuts in the RBF-detrended flat image. In most cases, there is a visible artefact in the top left corner of the RBF-detrended flats, which is typical of RBF interpolation, but this did not impact any of the stars studied in this work.

\subsubsection{Image Calibration}

All science images were calibrated in five steps. First, the master bias was subtracted. Next, the master dark was subtracted, correcting for the difference in exposure times between the science and dark images. The resulting image was then divided by the relevant corrected master flat frame for that night. Fourth, a bad pixel correction was applied, in which each bad pixel value was replaced with the median of its (up to) eight surrounding good pixels.

Finally, for each calibrated science image, a background map was produced. This was generated by masking the centre of the cluster and any remaining stars in the image and fitting the remaining region. Star detection and background fitting were carried out using \textsc{photutils} \citep{larry_bradley_2023_8248020}. The background maps were subtracted from each of the calibrated images. This process, combined with the flat-field correction, minimized the impact of scattered light both during twilight flats and night-time science observations.

\subsubsection{Astrometry}

We derived an initial World Coordinate System (WCS) solution for each calibrated image by extracting sources with \textsc{sep} \citep{bertin_sextractor_1996, barbary_sep_2016} and providing the resulting source table to \textsc{astrometry.net} \citep{lang_astrometrynet_2010}. To model image distortion, we used a Simple Image Polynomial (SIP; \citealt{2005ASPC..347..491S}) of degree 3.

To refine the initial \textsc{astrometry.net} fit, we re-fitted the source positions against a catalogue of stars from \textit{Gaia} DR3. The catalogued positions of the stars were adjusted by propagating their proper motions from the J2016.0 epoch to the time of the observations. We selected stars with magnitudes in the range $10<G<16$. Stars with neighbours brighter than three magnitudes dimmer ($\Delta G < 3$) within 22.5 arcseconds (50 binned pixels) were excluded to avoid centroid contamination. After applying these criteria, 110 \textit{Gaia} stars in the M3 field were available for use in the astrometric fit.

The pixel coordinates of stars extracted by \textsc{sep} were transformed into equatorial coordinates using the initial WCS solution and then cross-matched with the \textit{Gaia} catalogue. Matches were refined by excluding those with catalogued-to-measured separations exceeding 10 pixels or zeropoint magnitudes deviating by more than 3$\sigma$ from the mean. A degree 3 polynomial was then re-fitted to the filtered subset of matched stars to produce the final astrometric fit. This method for deriving the WCS solution consistently outperformed using \textsc{astrometry.net} alone. To quantify this improvement, we analysed the images obtained on 2023 December 13 (the first night of observations). For these images, between 77 and 98 matched stars per frame were used in the final fits. Across all images, the mean and standard deviation of the median residuals between measured stellar positions and those predicted by the fitted WCS were $0.088\pm0.009$\,arcsec using \textsc{astrometry.net} alone, and $0.061\pm0.007$\,arcsec using the improved method.

Additionally, at this stage in the pipeline, the non-variable (according to \textit{Gaia} DR3), uncontaminated matched stars were used to measure a final colour-dependent zeropoint magnitude of the whole image by applying a linear fit to the zeropoint (catalogued minus instrumental) magnitudes of the individual stars in $G$ against their $\textit{G}_\text{BP} - \textit{G}_\text{RP}$ colours.

\subsubsection{Aperture Photometry}\label{sec:photometry}

To facilitate the detection of transient objects and leverage existing information from \textit{Gaia}, we combined a catalogue-driven and a non-catalogue-driven approach. Following image calibration, the sources were extracted with \textsc{sep}. The signal-to-noise ratio (SNR) of each source in each image was calculated using the measured flux and the image-wide background RMS. Sources with SNR\,$<$\,5 were rejected. To avoid blended and non-stellar sources, we rejected sources with ellipticity greater than 0.5 measured by \textsc{sep}. We measured the half-flux diameters (HFDs) of the remaining sources and rejected images with median HFDs greater than 4~arcsec.

Each extracted source was cross-matched with the brightest \textit{Gaia} source within 0.5~arcsec. Unmatched sources were retained for manual inspection as potential transients. However, upon review, all unmatched sources were identified as galaxies, asteroids, or camera defects.

Given the very high density of stars within the cluster, we approximated the level of contamination from other sources in the image by integrating the predicted contaminating flux over the photometric aperture of neighbouring \textit{Gaia} sources given their magnitudes, colours and the colour-dependent zeropoint of the image, assuming gaussian-shaped point spread functions (PSFs) with the measured median HFD. Sources with a predicted flux leakage of more than 1 per cent from other stars into the aperture were flagged as contaminated in that image. The contamination status of a source thus varied with the HFD of each image. In practice, the actual PSFs exhibited axial asymmetry and broader tails than a gaussian, meaning this approach provided only an approximate indication of contamination.

Initially, we explored a difference imaging method using \textsc{hotpants} \citep{2015ascl.soft04004B}, similar to the approach taken by \cite{mccormac_search_2014} in their survey of M71. However, TWIST’s significant susceptibility to wind shake produced complex PSFs that the software could not effectively handle. Very large photometric apertures would have been needed, increasing the noise and the impact of contamination, negating the benefit of the approach. Consequently, we employed simple aperture photometry for our analysis.

We tested a range of aperture radii from 2 to 15 (binned) pixels and measured the resulting SNR in the light curves. Larger apertures were beneficial for brighter stars, while smaller apertures yielded better results for dimmer stars and in bright conditions. Ultimately, we adopted a 5-pixel aperture radius for the entire survey, which was optimal or near-optimal for stars with $14<G<19$. This consistent aperture size also ensured homogeneous data processing.

The flux error for each star in each image was calculated based on the background RMS and Poisson shot noise. The error model did not account for other noise sources, such as atmospheric scintillation and HFD-dependent contamination.

In total, 4,199 stars were detected in at least one valid image, 2,694 in at least 1 per cent of valid images, 1,038 in at least 50 per cent and 372 in at least 99 per cent. Stars detected in fewer than 50 per cent of images were generally dimmer than around $G=18.7$. The variation in the number of detections primarily reflects changes in seeing, sky transparency, and sky brightness, which affected the limiting magnitude between exposures.

We converted the fluxes of all the matched sources to $G$ magnitudes using the colour-dependent zeropoint magnitude of each image. This conversion does not account for the star’s full spectral energy distribution, intrinsic colour variations, or non-linearity in the zeropoint-colour relation, but it offers a practical approach for handling large stellar samples where a precise absolute photometric calibration is not required.

Additionally, we ran forced photometry with the same 5-pixel aperture size on the proper motion-propagated positions of all the known RRL variables within the field from \cite{clement_variable_2001} after cross-matching with \textit{Gaia} DR3. We excluded stars for which we could not find a cross-match in \textit{Gaia} DR3, specifically V187, V244, V248, V251, and V265, as \textit{Gaia} photometric parameters were required for the subsequent analysis. These excluded sources lie 11--32\,arcsec from the cluster centre in regions of severe crowding. We also excluded V297, identified as a foreground non-RRL variable \citep{bhardwaj_near-infrared_2020}. Finally, four variables (V123, V205, V206, and V299) fell outside the survey field of view, leaving 231 stars in the final sample.

\section{Periodic Variable Analysis}
\label{sec:freq_analysis}

This section describes the search for and analysis of periodic variables in the direction of the cluster.

\subsection{Periodicity Search}

On all of the corrected light curves, we performed a Lomb-Scargle \citep[LS -- ][]{lomb_least-squares_1976, scargle_studies_1982} periodogram using the implementation of \textsc{astropy} \citep{astropy:2013, astropy:2018, astropy:2022}. We followed the prescription of \cite{vanderplas_understanding_2018} and set a frequency grid with spacing, $\Delta f=1/n_oT=0.0005$\,$\mathrm{d}^{-1}$ where $T$ is the length of the survey, $\sim$200\,d, and $n_o=10$ is the number of test frequencies per peak, set to the same value as in \cite{richards_construction_2012} in their work on classifying variable stars. We set a minimum frequency of 0.0005\,$\mathrm{d}^{-1}$ and a maximum frequency of 60\,$\mathrm{d}^{-1}$, allowing the detection of SXP stars' pulsations, including higher harmonics. In most cases, some residual signal remained from the detrending with frequencies of integer multiples of 1\,$\mathrm{d}^{-1}$. We masked out frequencies within 5 per cent of 1\,$\mathrm{d}^{-1}$ and within 2 per cent of 2, 3 and 4\,$\mathrm{d}^{-1}$. This resulted in masking out the primary frequencies of some RRL stars. However, the large amplitudes of their non-sinusoidal pulsations allowed detection at higher harmonics. We additionally masked frequencies within 5 per cent of the lunar synodic frequency, $1/29.53$\,$\mathrm{d}^{-1}$.

\begin{figure*}
    \begin{center}
    \includegraphics[width=175mm]{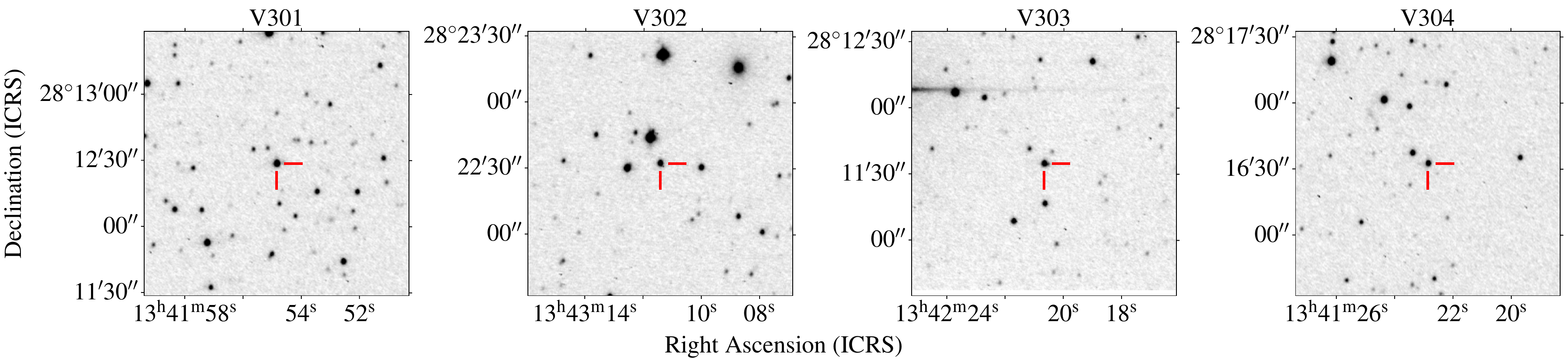}
    \end{center}
    \caption{Finding charts showing the locations of the newly discovered SXP-type variables V301--304. Each chart is 2$\times$2\,arcmin across. The images were created by aligning with \textsc{spalipy} \protect\citep{lyman_spalipy_2021} and stacking 360 calibrated frames (totalling one hour of exposure) with measured half-flux diameters less than 2\,arcsec taken during dark sky environments. North is up and East is to the left.}
    \label{fig:finding}
\end{figure*}

We visually inspected all stars with a full amplitude at their peak-power frequency of greater than 0.01 mag and rejected obviously spurious signals. The remaining periodic variables were all matched to currently known variable stars in the catalogue of \citet{clement_variable_2001}, with the exception of four. We classified these stars as SXP variables and designate them V301--304, from brightest to dimmest in $G$, following on from the discovery of the long-period variable V300 by \cite{siegel_swift_2015}. Finding charts for these stars can be found in Fig.~\ref{fig:finding}.

\subsection{SX Phoenicis Variables}\label{sec:sx}
\label{subsec:sxp}

SXP variables form a metal-poor subclass of $\delta$~Scuti variables. They lie on or above the zero-age main sequence in the instability strip of the Hertzsprung-Russell (HR) diagram with typically A or F spectral types. They follow a period-luminosity relation (e.g. \citealt{mcnamara_delta_2011}, \citealt{cohen_sx_2012}) which has been used to estimate distances to globular clusters and nearby galaxies. Compared to the more metal-rich $\delta$~Scuti variables, SXP stars are less massive, less luminous, and have, on average, shorter pulsation periods.

In this section, we demonstrate that the variables V301--304 are cluster members, that they are members of the SXP class, and, using their pulsation frequencies, provide an independent estimate of the distance to M3.

We cross-matched the four new SXP variables with their corresponding \textit{Gaia} DR3 sources and show their associated source IDs and astrometric and photometric properties in Table~\ref{tab:sx-phe-gaia}. Each has a parallax consistent with zero to 2$\sigma$, indicating they are not nearby field stars. Additionally, V301--304 have angular separations from the cluster centre of 10.8, 13.2, 11.2, and 12.3 arcmin, all well within M3's 28.7 arcmin tidal radius \citep{harris_catalog_1996}, indicating that they are coincident with the cluster.

\begin{table*}
	\centering
	\caption{Relevant photometric and astrometric properties of the four newly discovered SXP type variables from \textit{Gaia} DR3. Positions and proper motions are with respect to the ICRS reference frame at the J2016.0 epoch. $\mu_\alpha \cos{\delta}$ is the proper motion in right ascension corrected for declination, $\mu_\delta$ is the proper motion in declination. The \textit{V} band magnitude is derived from the sources' \textit{G} and $\textit{G}_\text{BP} - \textit{G}_\text{RP}$ magnitudes using the relation from Chapter 5.5.1 of version 1.3 of the \textit{Gaia} DR3 documentation \protect\citep{busso_gaia_2022}.}
	\label{tab:sx-phe-gaia}
	\begin{tabular*}{\linewidth}{@{\extracolsep{\fill}} llcccccccc }
		\hline
		  \textbf{V} & \textbf{\textit{Gaia} DR3 ID} & \textbf{RA} & \textbf{Dec} & \boldmath{$\mu_\alpha\cos{\delta}$} & \boldmath{$\mu_\delta$} & \textbf{Parallax} & \textbf{\textit{G}} & \textbf{$\textit{G}_\text{BP} - \textit{G}_\text{RP}$} & \textbf{\textit{V}} \\
           & & \textbf{(ICRS)} & \textbf{(ICRS)} & \textbf{(mas/yr)} & \textbf{(mas/yr)} & \textbf{(mas)} & \textbf{(mag)} & \textbf{(mag)} & \textbf{(mag)}  \\
		\hline
            301 & 1454778304460743936 & 13$^{\mathrm{h}}$41$^{\mathrm{m}}$54\fs85 & +28$^{\mathrm{\circ}}$12\arcmin28\farcs6 & -0.24 $\pm$ 0.11 & -2.51 $\pm$ 0.07 & 0.05 $\pm$ 0.10 & 17.61 & 0.39 & 17.66 $\pm$ 0.03\\
            302 & 1454828912055208192 & 13$^{\mathrm{h}}$43$^{\mathrm{m}}$11\fs42 & +28$^{\mathrm{\circ}}$22\arcmin32\farcs2 & -0.38 $\pm$ 0.16 & -2.77 $\pm$ 0.10 & 0.18 $\pm$ 0.13 & 18.20 & 0.32 & 18.24 $\pm$ 0.03 \\
            303 & 1454730746283291264 & 13$^{\mathrm{h}}$42$^{\mathrm{m}}$20\fs67 & +28$^{\mathrm{\circ}}$11\arcmin34\farcs8 & -0.19 $\pm$ 0.16 & -2.69 $\pm$ 0.09 & 0.01 $\pm$ 0.14 & 18.26 & 0.31 & 18.30 $\pm$ 0.03 \\
            304 & 1454782595128645376 & 13$^{\mathrm{h}}$41$^{\mathrm{m}}$22\fs86 & +28$^{\mathrm{\circ}}$16\arcmin32\farcs5 & -0.18 $\pm$ 0.17 & -2.56 $\pm$ 0.11 & -0.10 $\pm$ 0.16 & 18.27 & 0.36 & 18.32 $\pm$ 0.03\\
		  \hline
	\end{tabular*}
\end{table*}

According to \citet{libralato_hubble_2022}, the absolute proper motion of M3 is $(\mu_\alpha\cos{\delta}, \mu_\delta) = (-0.261 \pm 0.055, -2.674 \pm 0.040)$\,mas/yr. To assess the consistency of each star's motion with that of the cluster, we computed the two-dimensional Mahalanobis distance between the star and cluster proper motions, treating the right ascension and declination components as statistically independent and combining the star and cluster uncertainties in quadrature.

Given that the squared Mahalanobis distance follows a $\chi^2$ distribution with two degrees of freedom, we calculated the probability ($p$-value) of observing larger differences in proper motion under the assumption that each star shares M3’s motion. For V301--304, we found $p=0.11,0.53,0.89$, and $0.56$ respectively. In all cases, these probabilities indicate that the stars' proper motions are statistically consistent with that of M3, given the measurement uncertainties.

On the basis that each star is coincident with the cluster within the sky-projected tidal radius, has a proper motion consistent with the cluster, is not nearby, and is located far from the galactic plane, away from large numbers of field stars, we verify the membership of M3 of variables V301--304.

\begin{figure}
	\includegraphics[width=8.5cm]{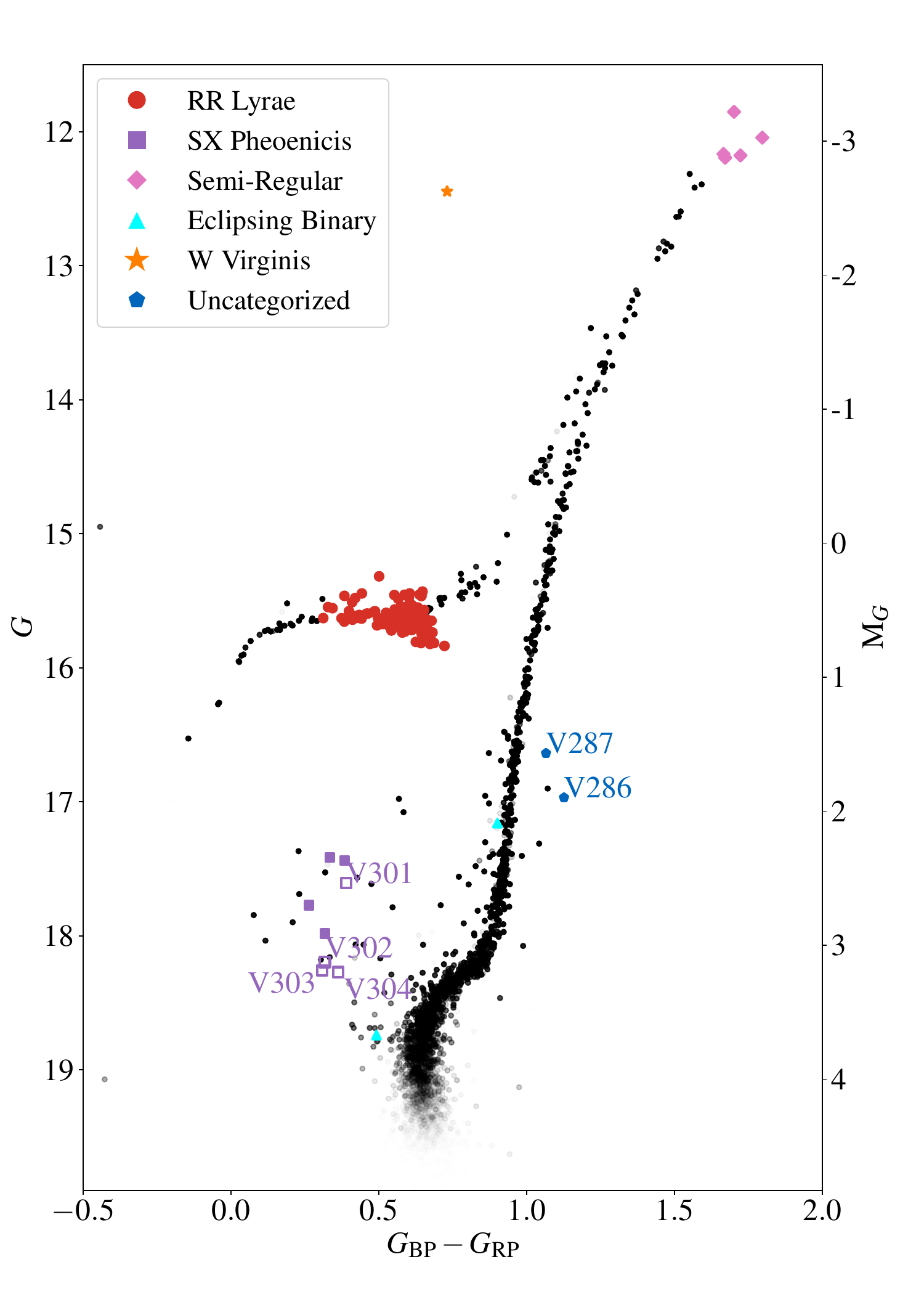}
    \caption{Colour-magnitude diagram of the \textit{Gaia} sources in the survey field with proper motions within three times the central velocity dispersion of the mean proper motion of the cluster (values taken from \protect\citealt{libralato_hubble_2022}), which we take to signify cluster membership in the absence of a more detailed analysis. M$_\textit{G}$ is the absolute \textit{G} magnitude assuming a distance modulus of $(m-M)_\textit{V}=15.07$ per \protect\cite{harris_catalog_1996} and zero colour excess, $E(\textit{V}-\textit{G})$. The locations of the newly discovered SXP-type variables are shown as open squares. Other variables are indicated as in Fig. \ref{fig:field}. The opacity of each point is given by the proportion of images in which its corresponding star is detected with at most 1 per cent contamination and at least at 5$\sigma$.}
    \label{fig:cmd}
\end{figure}

With their cluster membership confirmed, we place the variables on the colour-magnitude diagram (CMD, shown in Fig. \ref{fig:cmd}), constructed from the \textit{Gaia} sources in the survey field with proper motions within three times the central velocity dispersion of M3's proper motion (taken from \citealt{libralato_hubble_2022}). V301--304 are located brighter and bluer than the main sequence turn-off in the blue straggler region, indicating that they are themselves blue straggler stars.

The four new variables have detected peaks in their LS periodograms significantly greater than the 0.1 per cent false alarm probability (FAP) threshold at frequencies that could have no apparent systematic, non-astrophysical cause. The frequencies of the highest amplitude oscillations of each star fall in the interval given by \citet{mcnamara_delta_2011} for SXP stars -- 7.9$-$29.5\,d$^{-1}$. V301 falls in the overlapping region of the period ranges for SXP and $\delta$~Scuti stars, but with a metallicity, [Fe/H], of $-2.04$\,dex, derived from \textit{Gaia} DR3's Astrophysical parameters inference system \citep[Apsis;][]{creevey_gaia_2023}, the SXP classification is preferred. Since the four stars are blue straggler cluster members with pulsation periods typical of SXP stars, we confirm their variable classification.

We searched for evidence of radial mode pulsation in the light curves of the SXP stars by considering the frequencies present. We fit an $n$-term Fourier sine series of the form:

\begin{equation}
    m=m_0+\frac{1}{2}\sum^{n}_{i=1}A_i\sin{(2\pi ift +\phi_i)}
    \label{eq:fourier_coefficients}
\end{equation}

to the light curve (in magnitudes) of each star at its peak LS power frequency, $f$. $m_0$ is the constant offset, $A_i$ are the (full) amplitudes, $\phi_i$ are the phases, and $t$ is barycentric time. The derived Fourier parameters are then given by $\phi_{ij}=j\phi_i-i\phi_j$ and $R_{ij}=A_i/A_j$.

\begin{figure*}
	\includegraphics[width=8.5cm]{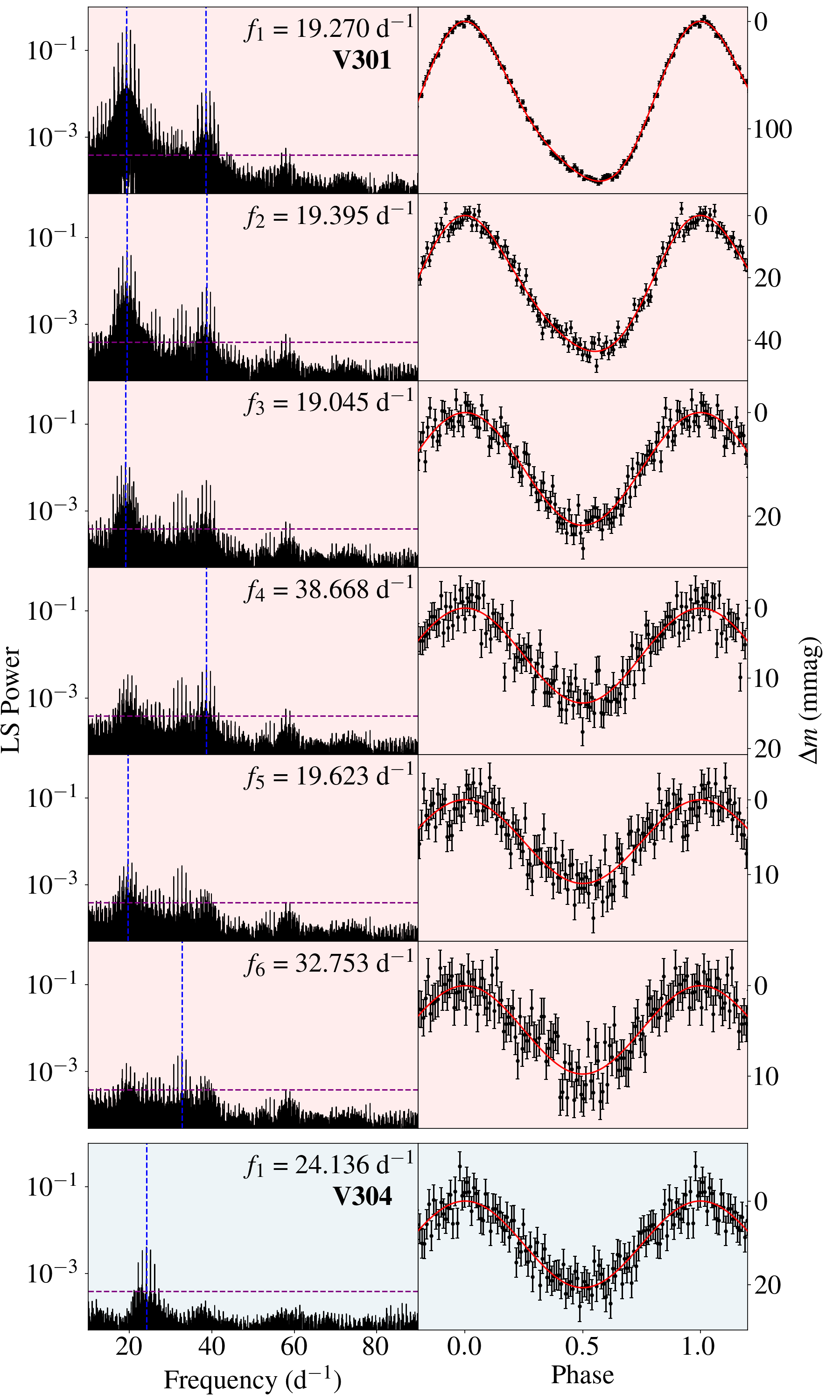}
        \includegraphics[width=8.5cm]{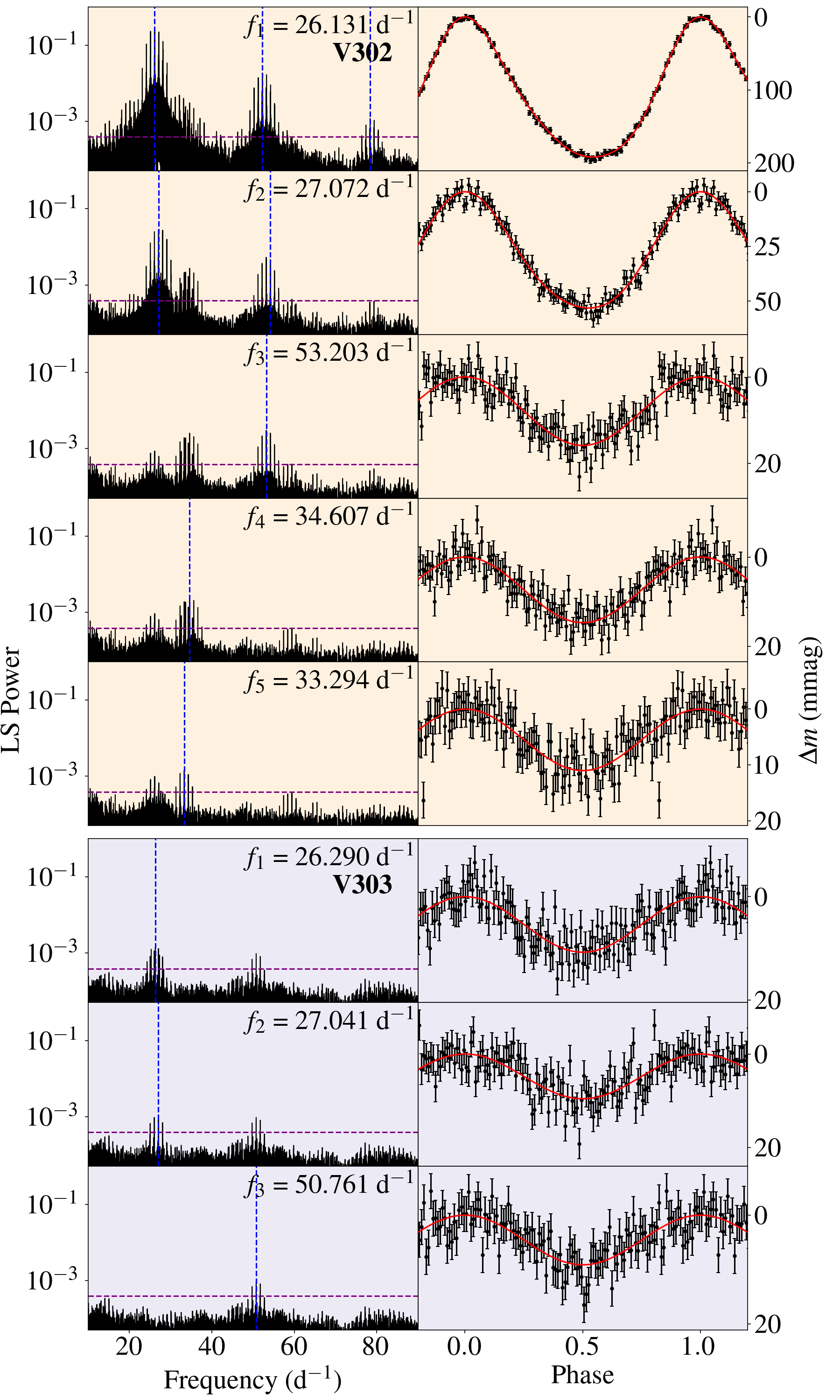}
    \caption{Lomb-Scargle periodograms (left panels of both plots) and phased light curves (right panels of both plots) of the newly discovered SX Phoenicis pulsators V301--304 (clockwise from the top-left), showing all of the extracted frequencies for each star. After each frequency is detected in the periodogram, the data are pre-whitened by subtracting a best-fit (multiple harmonic where possible) sinusoidal model of that frequency. The periodogram is recalculated after each pre-whitening step to extract additional frequencies. Vertical dashed lines in the periodograms mark the extracted frequencies, and in cases of multiple harmonic fits, the higher-order harmonics are also marked with dashed vertical lines. The horizontal dashed lines in the periodograms represent the 0.1 per cent false-alarm probability threshold. The phased light curves (right) show the data folded over each extracted frequency, with the best-fit models overlaid in red. For visibility, each light curve is binned by phase into 100 bins because the uncertainty of each measurement is, in most cases, considerably larger than the amplitude of the oscillation. Each light curve is normalized to the peak brightness of the fit.}
    \label{fig:301}
\end{figure*}

We performed an MCMC fit using \textsc{emcee} \citep{foreman-mackey_emcee_2013}, using unbounded uniform priors and allowing the frequency to vary as a free parameter. We used as many sine terms in the fit as had an amplitude detected with SNR of at least 2. We then pre-whitened the light curves by the Fourier fit and computed new LS periodograms. We repeated this process until the highest remaining peak was due to residual signal from the largest amplitude pulsation (V301--302) or there were no remaining peaks above the 0.1 per cent FAP threshold (V303--304). The Fourier fits and periodograms are shown in Fig. \ref{fig:301}, and the extracted Fourier parameters are shown in Table \ref{tab:sxp_fourier}.

\begin{table*}
	\centering
	\caption{Fourier properties of the newly discovered SXP stars. Magnitudes are in the instrumental system of TWIST. Amplitudes are the full peak-to-peak variations. Where only a single harmonic is detected, the total amplitude is identical to $A_1$.}
	\label{tab:sxp_fourier}
	\begin{tabular*}{\linewidth}{@{\extracolsep{\fill}} lllllllll }
		\hline
		  \textbf{V} & \textbf{Label} & \textbf{Frequency (d\boldmath{$^{-1}$})} & \textbf{Total amplitude (mag)} & \boldmath{$A_1$} \textbf{(mag)} & \boldmath{$\phi_{21}$} \textbf{(rad)} & \boldmath{$R_{21}$} & \boldmath{$\phi_{31}$} \textbf{(rad)} & \boldmath{$R_{31}$} \\
		\hline
            \textbf{301} & $f_1$ & 19.27057 $\pm$ 0.00001 & 0.1487 $\pm$ 0.0006 & 0.1452 $\pm$ 0.0006 & 2.30 $\pm$ 0.03 & 0.155 $\pm$ 0.004 &   &   \\
             & $f_2$ & 19.39537 $\pm$ 0.00004 & 0.0436 $\pm$ 0.0006 & 0.0430 $\pm$ 0.0006 & 2.35 $\pm$ 0.12 & 0.112 $\pm$ 0.013 &  &   \\
             & $f_3$ & 19.04511 $\pm$ 0.00009 &   & 0.0217 $\pm$ 0.0006 &   &   &  &   \\
             & $f_4$ & 38.66771 $\pm$ 0.00016 &   & 0.0135 $\pm$ 0.0006 &   &   &  &   \\
             & $f_5$ & 19.62343 $\pm$ 0.00016 &   & 0.0112 $\pm$ 0.0006 &   &   &  &   \\
             & $f_6$ & 32.75288 $\pm$ 0.00021 &   & 0.0098 $\pm$ 0.0006 &   &   &  &   \\
            \hline
            \textbf{302} & $f_1$ & 26.13173 $\pm$ 0.00002 & 0.1918 $\pm$ 0.0013 & 0.1865 $\pm$ 0.0009 & 2.04 $\pm$ 0.03 & 0.178 $\pm$ 0.005 & 4.16 $\pm$ 0.15 & 0.034 $\pm$ 0.005 \\ 
            & $f_2$ & 27.07126 $\pm$ 0.00006 & 0.0531 $\pm$ 0.0009 & 0.0530 $\pm$ 0.0009 & 1.87 $\pm$ 0.16 & 0.113 $\pm$ 0.018 &  &   \\
            & $f_3$ & 53.20279 $\pm$ 0.00019 &   & 0.0158 $\pm$ 0.0009 &   &   &  &   \\
            & $f_4$ & 34.60693 $\pm$ 0.00020 &   & 0.0147 $\pm$ 0.0009 &   &   &  &   \\
            & $f_5$ & 33.29425 $\pm$ 0.00028 &   & 0.0110 $\pm$ 0.0009 &   &   &  &   \\
	   \hline
           \textbf{303} & $f_1$ & 26.28969 $\pm$ 0.00026 &   & 0.0107 $\pm$ 0.0010 &   &   &  &   \\
           & $f_2$ & 27.04032 $\pm$ 0.00031 &   & 0.0096 $\pm$ 0.0009 &   &   &  &   \\
           & $f_3$ & 50.76065 $\pm$ 0.00039 &   & 0.0091 $\pm$ 0.0011 &   &   &  &   \\
        \hline
        \textbf{304} & $f_1$ & 24.13592 $\pm$ 0.00016 &   & 0.0207 $\pm$ 0.0010 &   &   &  &   \\
        \hline
	\end{tabular*}
\end{table*}

In all periodograms, we observe frequency aliases at multiples of 1\,d$^{-1}$ separations from the highest peaks, which can be attributed to the observing window function. In many cases, this makes the identification of the correct peak uncertain. The 10-second survey cadence improves the definition of the true pulsation peak by reducing high-frequency alias contamination and providing denser sampling of each cycle, although daily 1\,d$^{-1}$ sidelobes from the window function remain unavoidable and complicate peak selection in several cases. With the exception of $f_1$ and $f_2$ of V301 and V302, whose higher harmonics constrain the correct peaks, the extracted frequencies may differ from the intrinsic frequencies by 1 or 2\,d$^{-1}$, which would be a much larger error than the fit uncertainties. This is particularly relevant to V303, which has three other peaks with at least 90 per cent of the LS power of the most significant peak.

From \citet{olech_cluster_2005}, we expect the period ratios of the first few adjacent radial modes to be in the vicinity of the interval from 0.77 to 0.85 (see their Fig. 6). We identified two pairs of frequencies with (inverse) ratios in this interval. Firstly, $f_4$ and $f_6$ of V301 have a ratio of 0.847, which matches the third to second overtone period ratio according to the models of \citet{olech_cluster_2005}. However, $f_4$ itself appears to be almost exactly equal to $f_1+f_2$, meaning it is not an independent mode (and especially not a radial mode) but instead arises from non-linear interactions between $f_1$ and $f_2$.

In contrast, $f_1$ and $f_5$ of V302 have a ratio of 0.785, which is characteristic of the fundamental and first overtone radial modes. Given that V302 also has a relatively large amplitude at $f_1$, we conclude that its largest amplitude variability is due to pulsation in the fundamental radial mode. Knowing this, we can use a period-luminosity relation for fundamental mode pulsators to estimate the distance modulus to V302. Using the relation from Chapter 5.5.1 of version 1.3 of the \textit{Gaia} DR3 documentation \citep{busso_gaia_2022}, we estimate from the \textit{Gaia} photometry the apparent $V$ magnitude of the variables (shown in Table \ref{tab:sx-phe-gaia}), and using equation 5b of \citet{mcnamara_delta_2011} for the metal-poor fundamental mode period-lumonisity relation, we calculate the predicted absolute magnitudes in the $V$ band. Propagating the statistical and systematic uncertainties, we estimate a distance modulus $(m-M)_\textit{V}$ to V302 (and therefore M3) of $15.02 \pm 0.10$, in excellent agreement with the literature value of 15.07 from the catalogue of \citet{harris_catalog_1996}. 

From \citet{vandenberg_constraints_2016}, we adopt a reddening of $E(\textit{B}-\textit{V}) = 0.013$ and assume the canonical diffuse–ISM value $R_\textit{V} = 3.1$ \citep{cardelli_relationship_1989}. This yields a \textit{V}-band extinction, $A_\textit{V}=0.040$ and a true distance modulus of $(m-M)_0 = 14.98 \pm 0.10$, which is in excellent agreement with the more modern true distance modulus estimate of $\mu=15.041\pm0.017$ (statistical) $\pm$ $0.036$ (systematic) of \citet{bhardwaj_near-infrared_2020}. The uncertainties in the slope and intercept of the period-luminosity relation dominate the uncertainty in the distance moduli. Although \citet{pych_cluster_2001} and \citet{olech_cluster_2005} have demonstrated that radial pulsations dominate for SXP stars even at low amplitudes, we refrain from assuming radial pulsations for estimating the distances to the other stars when evidence from the frequency ratios is not available.

We note that variables V301 and V302 are flagged as variable stars in \textit{Gaia} DR3 and are broadly classified as belonging to one of the following types: $\gamma$ Doradus, $\delta$ Scuti, or SX Phoenicis by the classifier of \citet{rimoldini_gaia_2023}. However, their classifier does not distinguish between these specific classes for these sources. Within the variability tables of \textit{Gaia} DR3 \citep{gaia_collaboration_vizier_2022}, a primary frequency is listed for V301 (19.27084\,day$^{-1}$, with no stated uncertainty), which is in agreement with the value derived in this work (19.27057\,$\pm$\,0.00001\,day$^{-1}$). No frequency is reported for V302. The newly identified variables V303 and V304 are not flagged as variable in \textit{Gaia} DR3, likely owing to their faintness and the low amplitude of their variability, demonstrating that the TWIST observations are sensitive to sources below the \textit{Gaia} variability detection threshold.

V301--304 are the first new SXP stars to be discovered in M3 since the discovery of six by \citet{hartman_bvi_2005}. The present study increases the number of known SXP variables in the cluster from nine to thirteen.

\subsection{RR Lyrae Variables}
\label{subsec:rrl}

RR Lyrae stars are low-mass metal-poor radial pulsators on the horizontal branch of the HR diagram in the classical instability strip. They are commonly found in globular clusters, and M3 hosts the most known RRL stars of any globular cluster in the Milky Way \citep{clement_variable_2001}.

\subsubsection{Period and Amplitude Determination}

Incidental to our search for new variability in the cluster, we ran forced photometry on 231 previously discovered RRL stars within the survey field with cross-matches to \textit{Gaia} sources. We studied these stars for evidence of period changes, the appearance or disappearance of pulsation modes, and Blazhko-like amplitude modulation. Given the non-sinusoidal shape of RRL pulsations, we used the phase dispersion minimization technique (PDM; \citealt{stellingwerf_period_1978}) to identify periodicity in the extracted light curves. Specifically, we used the \textsc{pdm2b} data-rich implementation, which attempts to measure period and amplitude changes.

For each star, we ran the PDM analysis in two different modes: first, without segmenting the data to obtain a best estimate of the mean period and amplitude, and second, segmenting the data into overlapping windows of 7,500 contiguous points, sliding each time by 500 points to measure period and amplitude modulation. Full phase coverage within the segmented intervals was essential for resolving time-dependent amplitude variations, and this was substantially aided by the 10-second cadence of the TWIST observations. In both the segmented and non-segmented cases, we tested frequencies within 0.02\,d$^{-1}$ of the highest peak of an initial LS search. Each star had around 106,000 data points passing basic quality checks -- less than four arcsec HFD and mean zeropoint magnitude greater than 22.7.

We derive the periods from the lowest $\theta$~statistic period of the whole-data PDM runs. We used bootstrap resampling (with replacement) to generate 500 simulated light curves and PDM period samples for each star. We derive the uncertainties in the periods from the standard deviation of the distribution of extracted periods.

Overall, the photometry was of sufficient quality such that 203 of the 231 RRL stars had one visually distinct peak in the PDM periodogram over the tested frequency range and an associated FAP\,$<$\,0.001 according to the statistic of \citet{schwarzenberg-czerny_correct_1997}. The periods of these stars are shown in Table \ref{tab:rr_period_amp}. The remaining 28 stars, located 9.5--116\,arcsec from the cluster centre, were strongly affected by crowding and blending, to the extent that reliable period determination was not possible.

\begin{table}
	\centering
	\caption{Best fitting periods and amplitudes for the 203 observed RRL variables. The Blazhko flag indicates a strong positive detection of Blazhko-like modulation \textit{in our data}. Amplitudes are in the instrumental system of TWIST. The full table is available as supplementary material in machine-readable format.}
	\label{tab:rr_period_amp}
	\begin{tabular}{lcccc}
		\hline
		\textbf{V} & \textbf{Class} & \textbf{Bl.} & \textbf{Period (d)} & \textbf{Amplitude (mag)}\\
		\hline
1 & \textit{ab} & N & 0.5205813 $\pm$ 0.0000014 & 1.13 $\pm$ 0.04 \\
3 & \textit{ab} & N & 0.558206 $\pm$ 0.000003 & 1.14 $\pm$ 0.04 \\
4n & \textit{ab} & N & 0.585065 $\pm$ 0.000005 & 1.03 $\pm$ 0.04 \\
4s & \textit{ab} & N & 0.593010 $\pm$ 0.000009 & 1.05 $\pm$ 0.04 \\
5 & \textit{ab} & Y & 0.505856 $\pm$ 0.000004 & 0.76 $\pm$ 0.03 \\
... & ... & ... & ... & ... \\
261 & \textit{c} & N & 0.444723 $\pm$ 0.000012 & 0.41 $\pm$ 0.02 \\
266 & \textit{c} & N & 0.342354 $\pm$ 0.000012 & 0.53 $\pm$ 0.03 \\
270n & \textit{ab} & N & 0.62593 $\pm$ 0.00002 & 0.57 $\pm$ 0.03 \\
270s & \textit{ab} & N & 0.690186 $\pm$ 0.000006 & 0.87 $\pm$ 0.03 \\
290 & \textit{d} & N & 0.240398 $\pm$ 0.000003 & 0.0387 $\pm$ 0.0019 \\
		\hline
	\end{tabular}
\end{table}

\begin{figure}
    \begin{center}
    \includegraphics[width=85mm]{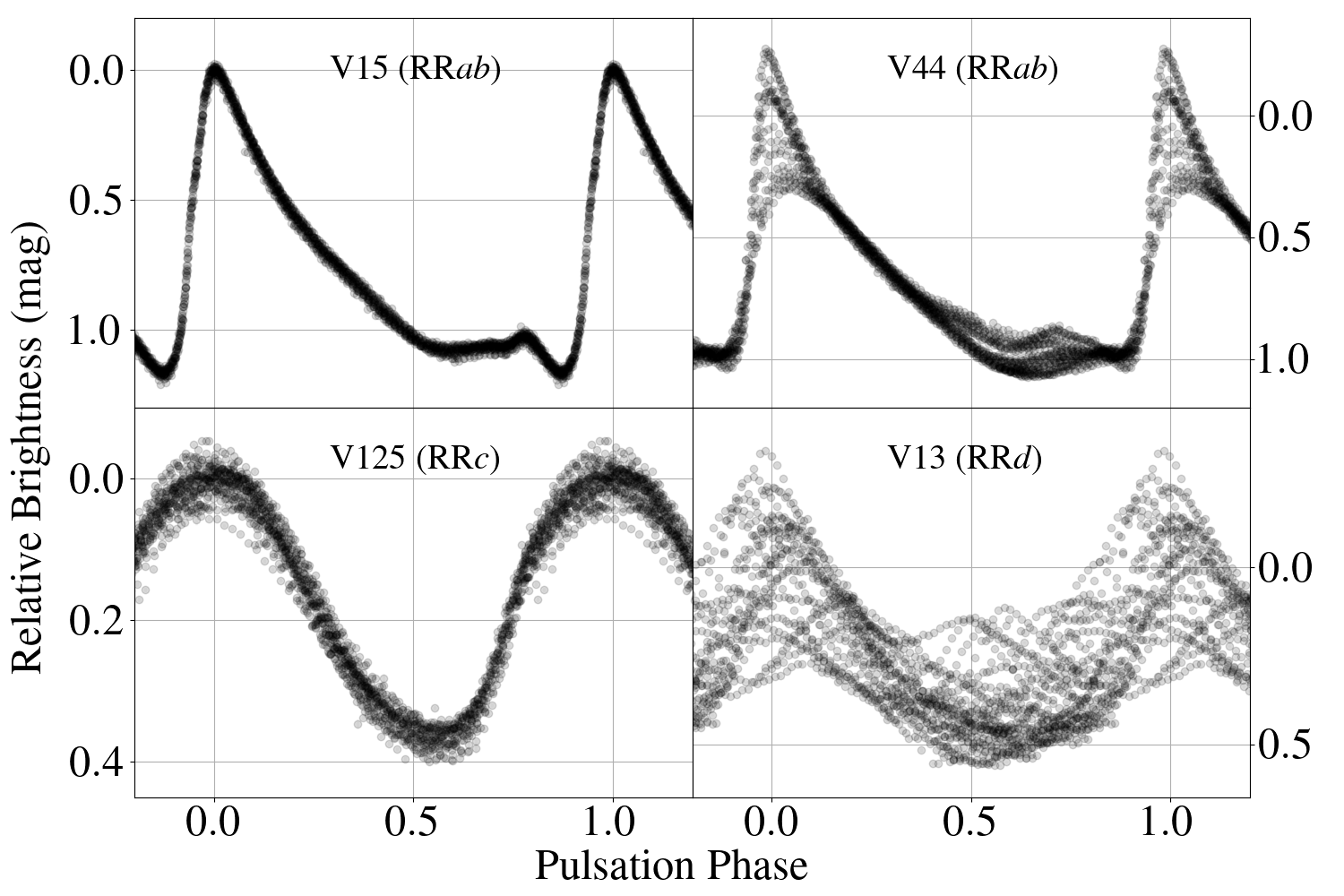}
    \end{center}
    \caption{Phased light curves for representative RR Lyrae stars observed with TWIST, illustrating the variety of pulsation behaviours. The upper panels show two fundamental-mode (RR\textit{ab}) stars, one stable and one showing Blazhko modulation. The lower panels show a first-overtone (RR\textit{c}) and a double-mode (RR\textit{d}) star. These examples include V13, V44, and V125, whose classifications are discussed in Section~\ref{sec:comparison_with_catalogued}. Phase is shown modulo one pulsation cycle, and relative brightness is plotted in magnitudes relative to the minimum magnitude of the whole-data \textsc{pdm2b} fit. The data are binned to 0.007\,d ($\sim$10-minute) bins prior to folding and the error bars are too small to see.}
    \label{fig:rr_puls}
\end{figure}

Since the majority of the variable stars were heavily contaminated in the dense field, to measure the amplitudes accurately, for each star we subtracted whatever constant flux was needed to make its zeropoint corrected flux-averaged magnitude identical to its catalogued \textit{Gaia} DR3 \textit{G} magnitude. This introduces a systematic uncertainty arising from the uncertainty in the absolute photometric calibration (the scatter about a linear fit to the data in Fig. \ref{fig:flux-vs-mag} -- 0.0413 mag), but permits amplitude measurements of highly contaminated stars, including 16 with no published amplitudes. The resulting systematic uncertainty is typically around $\sim$3-5 per cent of the amplitude in magnitudes and is the dominant source of uncertainty. We were able to measure the amplitudes of 198 of the stars with firmly detected periods; the remaining five have periods very close to one half or one third of a day, which either precluded complete phase coverage or artificially inflated the extracted amplitude owing to residual, nightly observing-induced periodicity with frequencies multiples of 1\,d$^{-1}$. Representative phase-folded light curves for the RR Lyrae stars with successfully determined periods and amplitudes are presented in Fig.~\ref{fig:rr_puls}.

\subsubsection{Comparison with Catalogued Properties}
\label{sec:comparison_with_catalogued}

\begin{figure}
    \begin{center}
    \includegraphics[width=85mm]{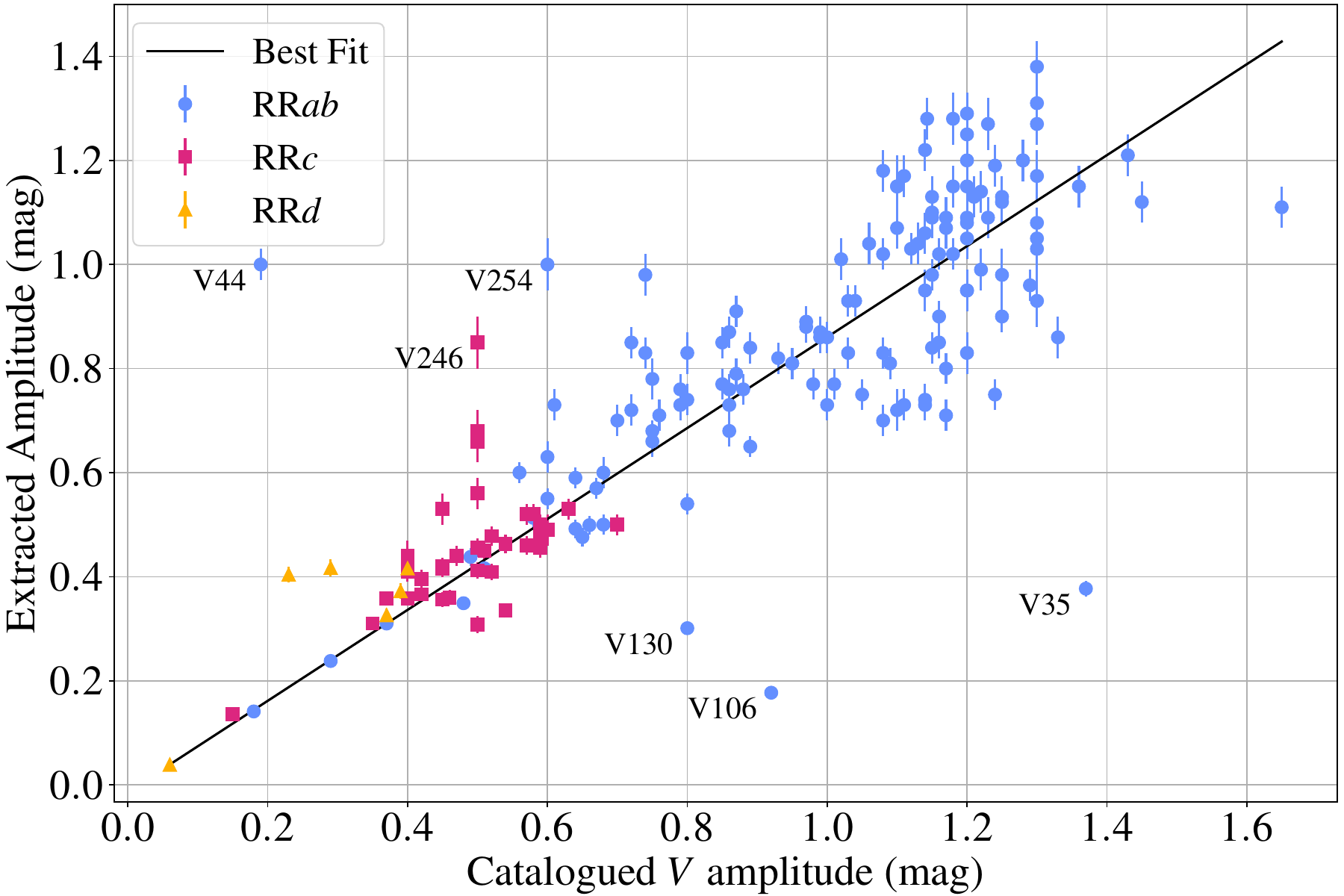}
    \end{center}
    \caption{Comparison between the catalogued \textit{V} amplitudes of the studied RRL stars from \protect\cite{clement_variable_2001}, with the extracted amplitudes of this study in the instrumental bandpass of TWIST. Outliers with residuals greater than 0.35 mag are labelled with their SAW IDs.}
    \label{fig:amp_comp}
\end{figure}

The extracted oscillation amplitudes are compared to those in the catalogue of \citet{clement_variable_2001} in Fig.~\ref{fig:amp_comp}. A 3$\sigma$ sigma-clipped linear fit to the data gives a slope of $\sim$87.3 per cent, a vertical intercept of $-0.013$\,mag, and negligible formal fit uncertainties. We observe smaller amplitudes in our instrumental band than the catalogued $V$ amplitudes, likely owing to TWIST having greater red coverage than the $V$ band. We do not expect a perfect one-to-one correspondence between the two sets of amplitudes, since intrinsic factors such as colour, temperature, and evolutionary phase within the instability strip introduce natural scatter. The observed correlation nonetheless confirms that the TWIST amplitudes are physically sensible and can therefore be used as a basis to estimate the $V$-band amplitudes of stars lacking published values, including the 16 stars for which we measured amplitudes but no $V$-band amplitude is available.

Of the six most significant outliers (labelled in Fig.~\ref{fig:amp_comp}), four of them (V35, V44, V106, and V130) show clear and significant Blazhko modulation, which contributes to the discrepancies. The remaining outliers, V246 and V254, are more likely to be statistical in nature, as their derived periods are consistent with the catalogued values. The discrepancies are likely due instead to larger measurement uncertainties caused by significant blending in the crowded central field.

Extracted periods generally matched the catalogued values of \citet{clement_variable_2001} to the precision with which they are quoted therein, and in all cases were within 1 per cent of the catalogued period. The variables V29, V41, V106, and V270n (all RR\textit{ab} type) stand out as deviating from the catalogue values by more than one minute. The period for V29 in \citet{clement_variable_2001} is listed as 0.47\,d on account of the discrepancy between the measurements of \citet{corwin_bv_2001} and \citet{benko_multicolour_2006}. Our measured period is inconsistent with \citet{corwin_bv_2001} by 52.7 seconds ($\sim$7.6 times our measurement uncertainty) but consistent with \citet{benko_multicolour_2006} to within 3.3\,seconds ($\sim$0.5$\sigma$). Similarly, the period of V270n is listed to only two significant figures, but our measurement is consistent with \citet{jurcsik_photometric_2017} to 6.9\,seconds ($\sim$4$\sigma$). For V41, our measured period is inconsistent with \citet{benko_multicolour_2006} by 72\,seconds ($\sim$21$\sigma$), but consistent with the value given in \citet{jurcsik_photometric_2017} to 5.2\,seconds ($\sim$1.5$\sigma$).

For V106, \citet{jurcsik_photometric_2017} report a fundamental pulsation period, $P_\mathrm{J}=0.54689~\mathrm{d}$, while we extract $P_\mathrm{our}=0.542424~\mathrm{d}$ -- $\sim$385\,s shorter. While period changes are expected in RRL stars, such a reduction on a timescale of roughly a decade is physically implausible \citep{jurcsik_long-term_2012}. We note that the difference between the two extracted frequencies, $f_\mathrm{our}-f_\mathrm{J}=0.0151~\mathrm{d}^{-1}$, is within a few per cent of the Blazhko frequency given in \citet{jurcsik_blazhko-type_2019}, $f_{\mathrm{Bl}} = 0.0157~\mathrm{d^{-1}}$, and therefore we interpret our extracted period as a Blazhko side peak rather than a genuine decrease in the fundamental pulsation period.

Following visual inspection of the phase-folded light curves, we agree with the RRL classifications given in \citet{clement_variable_2001}, except for V44 and V125, both of which they label as RR\textit{d}-type, and which we classify as RR\textit{ab} and RR\textit{c} respectively. In the case of V44, we agree with the reclassification of \citet{jurcsik_blazhko-type_2019} on the basis that no multiple periodicity is seen and that the characteristic sawtooth shape of an RR\textit{ab} star is present. 

In the case of V125, we disagree with its RR\textit{d} classification in the catalogue, which is taken from the work of \citet{jurcsik_overtone_2015}, and find no reclassification in any study since then. We note that \citet{bhardwaj_near-infrared_2020} observed V125 as part of their near-infrared survey of M3's RR Lyrae population in 2019 May and retained the multi-mode classification. In our data, only a singular periodicity is visible in the light curve, with the characteristic sinusoidal shape of an RR\textit{c} pulsator. We searched for evidence of fundamental-mode pulsation in the LS periodogram of its first overtone-mode prewhitened light curve at the expected frequency given in \citet{jurcsik_overtone_2015}, but we found none. Additionally, in the same reference, a second-overtone mode frequency is tenuously provided for the star V13, which, after pre-whitening by the fundamental and first-overtone modes, also cannot be detected in our data. This is despite the densely sampled and high-precision photometry. Thus, we consider that the fundamental mode of V125 and the second overtone mode of V13 (if ever present) have disappeared. Phase-folded light curves for V13, V44, and V125 are all shown in Fig.~\ref{fig:rr_puls}.

\subsubsection{Blazhko Modulation}

First described by \citet{blazko_mitteilung_1907}, the quasi-periodic phase and amplitude modulation of variable pulsators is known as the Blazhko effect. This effect, although poorly understood, has been observed in around half of the RR\textit{ab} stars in M3 \citep{jurcsik_long-term_2012}, which is a typical fraction for Oosterhof Type I clusters (e.g. \citealt{jurcsik_long-term_2011, smolec_rr_2017}). With our limited photometric precision for contaminated stars, we confidently recovered 53 stars showing visible, unambiguous amplitude modulation. In most cases, \citet{jurcsik_long-term_2012} lists these stars as positively identified modulators, with the exception of the RR\textit{c} stars V126 and V203, which were later identified by \citet{jurcsik_overtone_2015}.

\begin{figure*}
    \begin{center}
    \includegraphics[width=168mm]{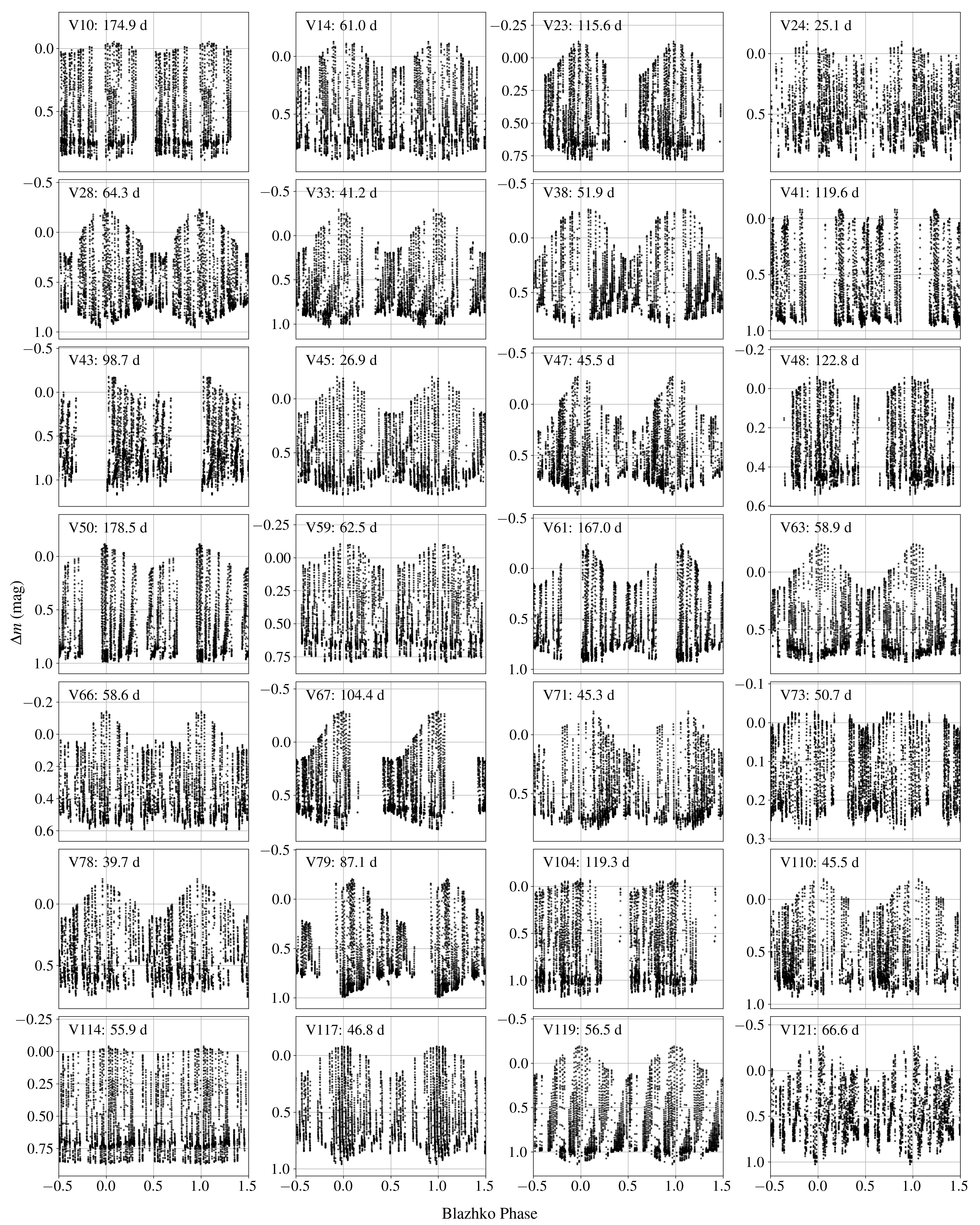}
    \end{center}
    \caption{Blazhko phase-folded light curves of the modulators listed in Table \ref{tab:rr_blazhko}. The folded light curves are centred on the maximum pulsation amplitude of the sine fit, and the data are binned to 0.007\,d ($\sim$10-minute) bins prior to folding. The SAW IDs and the best-fit modulation periods are shown in each plot.}
    \label{fig:blazhko}
\end{figure*}

To characterize the modulation of the Blazhko stars, we used the results of the sliding-window \textsc{pdm2b} analysis, which returns a period and amplitude for one 7500 data point window of the light curve at a time, sliding by every 500 points between measurements. We rejected measurements that spanned a range of more than 15 days to ensure the measurements better describe the instantaneous properties of the light curve. We additionally rejected data points for which the largest gap in the pulsation period-folded light curve was greater than 0.05 to reduce the probability of spuriously low amplitude measurements arising from incomplete phase coverage. We were not generally able to measure period modulation, and hence, we omit any such analysis. We fit the amplitude modulations each to a single harmonic sine wave and visually inspected the fits. This resulted in good fits for 28 stars, whose Blazhko phase-folded light curves are shown in Fig. \ref{fig:blazhko} and whose fit parameters are listed in Table \ref{tab:rr_blazhko}. For the other stars, either there were too few amplitude measurements on account of the phase coverage requirement, the modulation period was too long relative to the survey baseline, or the fits did not convincingly pass visual inspection. In the case of V35, the very large amplitude modulation ($\sim0.75$\,mag) could not be approximated as sinusoidal.

\begin{table}
\centering
	\caption{Fit parameters for a sinusoidal fit to the amplitude modulation of the constrained Blazhko stars. The uncertainties given are the 1$\sigma$ scatter in the MCMC fits. The modulation amplitude is the full peak-to-peak range of pulsation amplitudes. For comparison, we show the 2012 modulation periods from \protect\cite{jurcsik_blazhko-type_2019} (labelled J19) for these stars. Multiple modulation periods are separated by forward slashes, and lower bounds are denoted with colons.}
	\label{tab:rr_blazhko}

\begin{tabular}{lccc}
\hline
\textbf{V} & \textbf{Mod. Period (d)} & \textbf{J19 Mod. Period (d)} & \textbf{Mod. Amp (mag)} \\
\hline
10 & 175 $\pm$ 10 & 191 & 0.120 $\pm$ 0.007 \\
14 & 61.0 $\pm$ 0.3 & 60.6 & 0.317 $\pm$ 0.006 \\
23 & 115.6 $\pm$ 0.6 & 125 & 0.405 $\pm$ 0.006 \\
24 & 25.1 $\pm$ 0.7 & 49.8 & 0.22 $\pm$ 0.02 \\
28 & 64.3 $\pm$ 0.3 & 65.9 & 0.623 $\pm$ 0.013 \\
33 & 41.18 $\pm$ 0.19 & 40.5/34.8: & 0.59 $\pm$ 0.02 \\
38 & 51.9 $\pm$ 0.4 & 51.8 & 0.60 $\pm$ 0.03 \\
41 & 120 $\pm$ 5 & 190 & 0.147 $\pm$ 0.010 \\
43 & 98.7 $\pm$ 1.0 & 103 & 0.387 $\pm$ 0.015 \\
45 & 26.88 $\pm$ 0.11 & 26.6 & 0.426 $\pm$ 0.017 \\
47 & 45.5 $\pm$ 0.3 & 69/53.3/44.6 & 0.50 $\pm$ 0.03 \\
48 & 122.8 $\pm$ 1.8 & 143 & 0.121 $\pm$ 0.005 \\
50 & 178 $\pm$ 12 & 178: & 0.204 $\pm$ 0.011 \\
59 & 62.5 $\pm$ 0.3 & 62.1 & 0.238 $\pm$ 0.008 \\
61 & 167 $\pm$ 4 & 134: & 0.505 $\pm$ 0.010 \\
63 & 58.9 $\pm$ 0.3 & 58.7 & 0.390 $\pm$ 0.009 \\
66 & 58.6 $\pm$ 0.5 & 65.9/49.3 & 0.283 $\pm$ 0.009 \\
67 & 104.4 $\pm$ 0.4 & 103 & 0.618 $\pm$ 0.005 \\
71 & 45.3 $\pm$ 0.2 & 46.7 & 0.355 $\pm$ 0.017 \\
73 & 50.7 $\pm$ 0.7 & 54.3 & 0.073 $\pm$ 0.005 \\
78 & 39.7 $\pm$ 0.2 & 39.5 & 0.387 $\pm$ 0.015 \\
79 & 87.1 $\pm$ 0.6 & 160/62.9/18.3: & 0.530 $\pm$ 0.019 \\
104 & 119 $\pm$ 2 & 110 & 0.131 $\pm$ 0.005 \\
110 & 45.50 $\pm$ 0.17 & 44.9 & 0.347 $\pm$ 0.012 \\
114 & 55.9 $\pm$ 0.3 & 54.5 & 0.078 $\pm$ 0.002 \\
117 & 46.8 $\pm$ 0.2 & 47.2 & 0.314 $\pm$ 0.008 \\
119 & 56.5 $\pm$ 0.4 & 56.9 & 0.439 $\pm$ 0.010 \\
121 & 66.6 $\pm$ 0.6 & 65.1 & 0.45 $\pm$ 0.02 \\
\hline

\end{tabular}
\end{table}

All of the Blazhko stars for which we could fit the amplitude modulation are of type RR\textit{ab} and have published modulation periods in \citet{jurcsik_blazhko-type_2019}, which are shown in Table \ref{tab:rr_blazhko}. Although no uncertainties are given for the periods in that reference, we are generally in agreement for the stars with periods less than around 100\,d. Although the longer period modulators may have evolved, the discrepancies are likely due to a lack of Blazhko phase coverage and an over-reliance on the not perfectly applicable sinusoidal model, resulting in systematic uncertainties that are not captured by the fit uncertainties. The modulators for which we detect a significantly different period from the literature values are V24, V66, and V79. The detected period for V24 is very close to half the literature value, and the light curve is similarly well-fit by a 50.4\,d period. Due to the period aliasing, we cannot confidently rule out either period, although the 50.4\,d-folded light curve visually appears to have a greater amplitude, making it more likely. Our observations of V66 and V79 are clearly incompatible with the periods reported by \citet{jurcsik_blazhko-type_2019}, indicating that their modulation periods have changed since that study. Such changes are consistent with the quasi-periodic nature of the Blazhko effect and have been observed in RR Lyr itself \citep{borgne_historical_2014}.

We note that \citet{skarka_blazhko_2020} introduced a morphological classification to the Blazhko-phase folded light curves in the \textit{I} band; however, we refrain from classifying the observed Blazhko modulators according to this system due to the significantly different bandpass used in the present survey.

\subsection{Unclassified Variables}
\label{subsec:unc}

Two of the variable stars recovered in our survey are the unclassified variables V286 and V287. These stars were initially identified by \citet{hartman_bvi_2005}, who had insufficient phase coverage to determine reliable periods or amplitudes. Both variables were later assessed as highly probable cluster members in the membership study by \citet{prudil_membership_2024}. Their positions on the colour-magnitude diagram are shown in Fig. \ref{fig:cmd}, to the right of the giant branch.

V287 was subsequently observed in the near-UV by \citet{siegel_swift_2015} using the Ultra-Violet and Optical Telescope (UVOT) on \textit{Swift}; however, no significant variability was detected in those observations. More recently, both V286 and V287 were included in the \textit{Gaia} DR3 epoch photometry. We applied LS periodograms to the $G$ band photometry, yielding most likely periods of 16.5\,$\pm$\,0.2\,d for V286 and 4.773\,$\pm$\,0.011\,d for V287. These uncertainties were estimated from the half-width half-maxima of the respective periodogram peaks. Although both peaks exceed the 0.1 per cent FAP threshold, visual inspection of the phase folded light curves does not reveal convincing singular periodicity in either star. 

\begin{figure}
    \begin{center}
    \includegraphics[width=85mm]{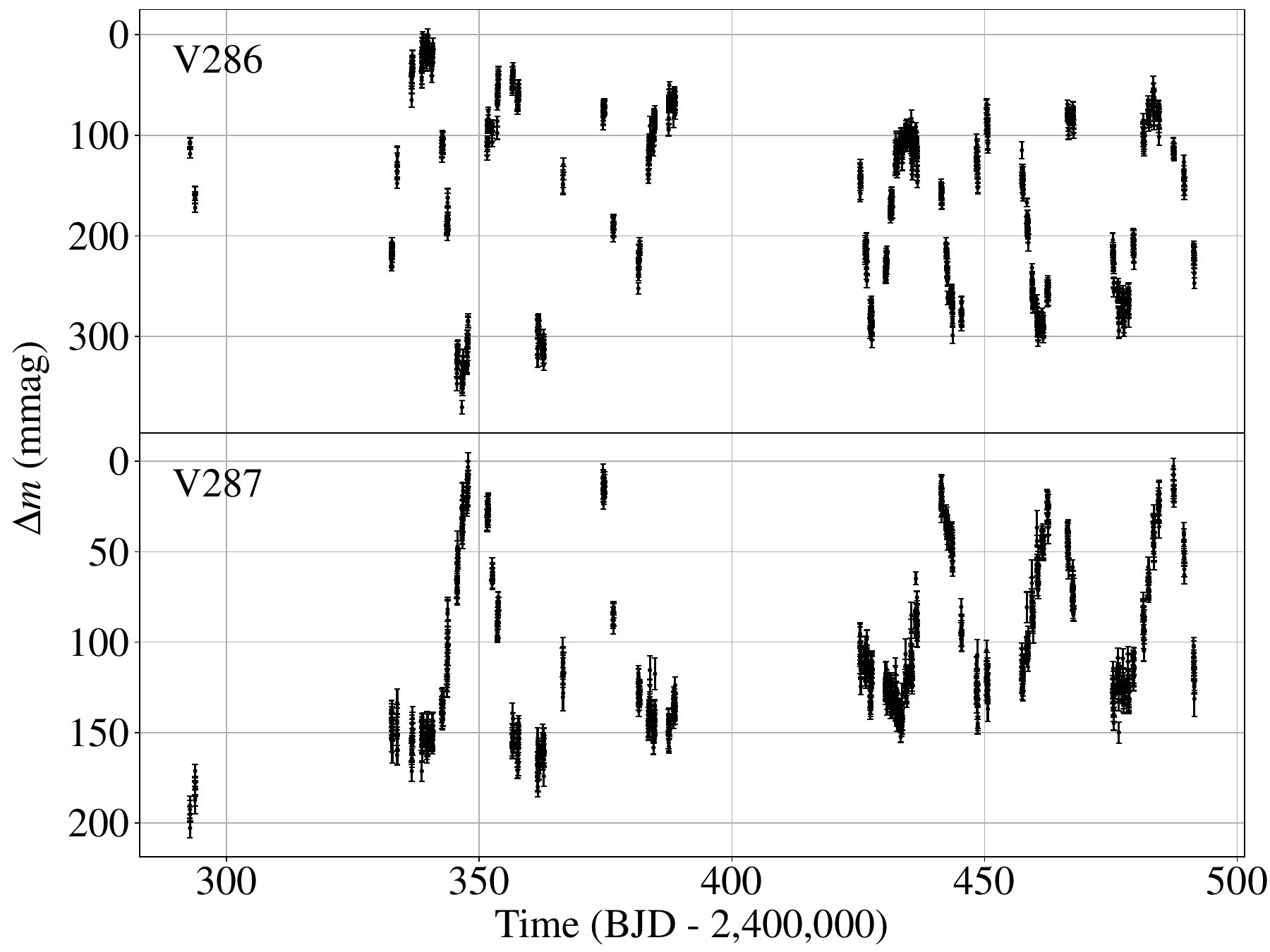}
    \end{center}
    \caption{Light curves of the unclassified variables V286 and V287 relative to their peak brightnesses in the instrumental bandpass of TWIST. The data are binned to 14.4\,min (0.01\,d) bins.}
    \label{fig:unclassified}
\end{figure}

Our observations (shown in Fig. \ref{fig:unclassified}) achieved full phase coverage over several cycles for each star. In both cases, the average variability is well fit by a single two-harmonic sine series, indicating a dominant single period. However, both light curves exhibit amplitude modulation on timescales longer than the survey baseline. As described in Section \ref{sec:sx}, we performed MCMC fits using a two-harmonic sine model. 

The fit results are summarized in Table \ref{tab:unclassified}. We extract a period for V286 that is consistent with the value from \textit{Gaia}. However, the period we measured for V287 does not match the \textit{Gaia} epoch photometry at all. Although we constrain the period and amplitude and have reliable estimates of the stars’ intrinsic brightnesses and colours, we are unable to classify either variable based solely on photometric data. The variability characteristics do not correspond to any known type in the General Catalogue of Variable Stars \citep{samus_general_2017}. Spectroscopic follow-up may be required to determine the nature of their variability. We suggest that, given their proximity on the colour-magnitude diagram, and the similarity of their periods and amplitudes, V286 and V287 are most likely of the same variable type.

\begin{table}
	\centering
	\caption{Properties of the observed unclassified variables. The amplitudes given are in the instrumental bandpass of TWIST and represent the average peak-to-peak amplitude over the observed window. The uncertainties are therefore concerning the \textit{average} amplitudes. The variations from one oscillation to the next are larger (see Fig. \ref{fig:unclassified}).}
	\label{tab:unclassified}
	\begin{tabular}{lccc}
		\hline
		\textbf{V} & \textbf{\textit{Gaia} DR3 ID} & \textbf{Period (d)} & \textbf{Amplitude (mag)}\\
		\hline
	286 & 1454782118391718528 & 16.3515 $\pm$ 0.0011 & 0.2463 $\pm$ 0.0005 \\
287 & 1454783320982452992 & 22.865 $\pm$ 0.003 & 0.1351 $\pm$ 0.0003 \\
		\hline
	\end{tabular}
\end{table}

\section{Non-Periodic Variability}
\label{sec:non-periodic}

\subsection{Microlensing}
\label{subsec:microlensing}

Globular clusters are, in theory, ideal locations to observe microlensing events against background stars \citep{paczynski_glob_1994}. Since the potential lenses have common and constrained distances and proper motions, resolving the lens at all is unnecessary. Additionally, knowing the lens distance and proper motion drastically reduces the complexity of often degenerate microlensing systems. If the background source's distance and proper motion can also be determined (or otherwise reasonably assumed), the problem of determining the lens mass reduces to measuring the Einstein crossing time \citep{gould_natural_2000}, which is trivial using photometry.

We considered the possibility of observing microlensing events from low-mass, sub-stellar lenses in our data by analytically predicting event rates under two scenarios -- firstly, with a lens anywhere in the cluster and the source far in the background (background lensing), and secondly, with a lens in the halo of the cluster near the tidal radius and the source in the centre of the cluster (self-lensing). 

\subsubsection{Self-lensing}

Due to dynamical mass segregation in globular clusters, lower-mass objects are expected to migrate toward the outer regions over time, while higher-mass stars sink towards the centre \citep{spitzer_equipartition_1969}. Hence, we assume for our model that potential lenses are located in a shell at a distance much greater than the cluster's stars. In this regime, the relative source-lens proper motion is dominated by the motion of the source in the cluster centre, so we do not consider the lens's motion. We take a microlensing "event" to be an alignment of a source and lens within the Einstein radius, $\theta_E$, which corresponds to a source amplification of around 1.34 times. The event rate, $\Gamma$, is then the area swept out by twice the Einstein radius over lens-source proper motion, multiplied by the surface density of source stars. This approach is inspired by that of \cite{kiroglu_gravitational_2022}. Formulaically, 

\begin{equation}
    \Gamma = \frac{v_\mathrm{d} N_\mathrm{s} N_\mathrm{l} \sqrt{G}}{\pi r^{3/2} c} \overline{\sqrt{M_\mathrm{l}}},
\end{equation}

where $N_\mathrm{s}$ is the number of sources, $N_\mathrm{l}$ is the number of lenses, $r$ is the radius of the shell of lenses, and $\overline{\sqrt{M_\mathrm{l}}}$ is the mean value of the square roots of the lens masses. $v_\mathrm{d}$ is the velocity dispersion of the cluster, which we use as a characteristic velocity of the sources in this approximation.

Using the central velocity dispersion (5.5\,km/s) from \citet{harris_catalog_1996}, the limiting radius (103\,pc), from \citet{trager_catalogue_1995}, the present-day mass of M3 (450,000\,M$_\odot$) from \citet{massari_multiple_2016}, and assuming an average stellar mass of 0.35\,M$_\odot$, we predict an \textit{inverse} event rate,

\begin{equation}
    \Gamma_\text{M3}^{-1} \approx 52300 \frac{N_\mathrm{s}}{N_\mathrm{l}} \frac{\sqrt{\mathrm{M_J}}}{\overline{\sqrt{M_\mathrm{l}}}} \left( \frac{r}{r_\mathrm{L}} \right)^{3/2} \text{yr},
\end{equation}

where $r_\mathrm{L}$ is the limiting radius. Hence, microlensing by sub-stellar lenses of the central stars would be extremely rare, on the order of once per tens of thousands of years, even if a large population of such lenses existed. Additionally, the sky-projected Einstein radius due to a Jupiter-mass lens at the limiting radius would be only $\sim$6.1\,R$_\odot$, meaning finite source effects would be present, reducing the maximum amplification and detectability.

\subsubsection{Background Lensing}

Similarly, we created an analytic model to predict the event rate in the scenario that a lens anywhere in the cluster amplifies a stellar source in the distant background. For simplicity, we assume the source is located infinitely far away and stationary, and that the relative source-lens proper motion is equal to the mean proper motion of the cluster.

In this case, the event rate is given by

\begin{equation}
    \Gamma = 2 N_\mathrm{l} \sigma_\mathrm{s} \mu \sqrt{\frac{4G}{c^2}\frac{1}{d}} \overline{\sqrt{M_\mathrm{l}}},
\end{equation}

where $\sigma_\mathrm{s}$ is the angular area density of sources, $\mu$ is the magnitude of the cluster proper motion, and $d$ is the distance to the cluster. Note that the product of the square-rooted terms (the mean Einstein radius) should be interpreted as an angle in radians, thus $\sigma_\mathrm{s}$ must be given in sr$^{-1}$ and $\mu$ in radians per unit time. Taking the stellar mass of the galactic halo beyond the cluster to be 3\,$\times$\,10$^8$\,M$_\odot$ -- the value for 10--45\,kpc from \citet{bland-hawthorn_galaxy_2016} -- and the mean stellar mass to again be 0.35\,M$_\odot$, we crudely estimate the source density behind the cluster to be 6.8\,$\times$\,10$^7$\,sr$^{-1}$, assuming a purely radial halo distribution and ignoring the difference in line of sight of the cluster between the sun and the galactic centre. We take the distance to the cluster to be 10.2\,kpc from \citet{bhardwaj_near-infrared_2020} and again use the cluster proper motion from \citet{libralato_hubble_2022}. This results in an inverse event rate of

\begin{equation}
    \Gamma_\text{M3}^{-1} \approx 3270\frac{N_\mathrm{stars}}{N_\mathrm{l}} \frac{\sqrt{\mathrm{M_J}}}{\overline{\sqrt{M_\mathrm{l}}}} \text{yr},
\end{equation}

where $N_\mathrm{stars}$ is the number of stars in the cluster and hence $N_\mathrm{l}/N_\mathrm{stars}$ is the number lenses per star. This result indicates that such microlensing events are still too rare to hope to observe: once every few millennia for the scenario of one Jupiter-mass lens per star. This does not even consider that most halo stars are too dim to be detected except by the largest space telescopes. If we consider instead microlensing by the cluster stars themselves, the event rate would be around an order of magnitude larger, but still far too rare to consider. 

\subsubsection{Microlensing Event-Rate Discussion}

The predicted rarity is perhaps not surprising. Past globular cluster microlensing searches have focused on the clusters M22 and 47 Tuc, which each have far richer stellar backgrounds than M3 -- the galactic bulge and the Small Magellanic Cloud, respectively, and these searches have borne little.

47 Tuc was observed over a baseline of almost five years, with a mean cadence between 2 and 3 days, in field SMC140 of the third iteration of the Optical Gravitational Lensing Experiment (OGLE-III, \citealt{wyrzykowski_ogle_2011}). No microlensing events were detected.

\citet{sahu_gravitational_2001} observed M22 with the \textit{Hubble Space Telescope} over a baseline of 113 days at a typical cadence of 3 days and reported one time-resolved and six time-unresolved microlensing events. However, all were later disputed \citep{sahu_reexamination_2002, anderson_probable_2003}. \citet{pietrukowicz_cluster_2005} reported the probable detection of a microlensing event in M22 with the 1-metre Swope telescope at Las Campanas Observatory, which was later confirmed by \citet{pietrukowicz_first_2011}, making it the first confirmed microlens in a globular cluster. To our knowledge, this remains the only confirmed example.

\citet{kiroglu_gravitational_2022} predicted microlensing event rates by M dwarf, white dwarf, neutron star, and black hole lenses for 47 Tuc and M22 using simulated motions of the cluster constituents, finding negligible event rates for self-lensing. Background lensing rates were dominated by white dwarfs and M dwarfs, and were at a rate of around a few per year. Given M3's much poorer stellar background, we would not expect to observe any microlensing events over the course of our observations due to background lensing. Self-lensing rates would likely occur at similar rates to those of M22 and 47 Tuc, and thus would also not be expected to be found in our data.

\subsubsection{Search for transient brightening}

Despite the predicted rarity of microlensing events, we searched for evidence of non-periodic transient brightenings in all the stars observed in the survey. In the reduced light curves, we checked for three consecutive points more than 5 times the median absolute deviation (in mag) brighter than the median brightness. Apart from obvious systematics, this identified three candidates with brightenings of 0.5-0.8 mag, lasting from around 7 to 30 minutes. Their \textit{Gaia} DR3 IDs are 1454876878250136576, 1454801733503276672, and 1454799087803284608. For these three, we verified that the brightenings were visible in both the raw flux and the inferred magnitudes and that the centroids remained stable over the event durations.

The three stars have $G_\text{BP}-G_\text{RP}$ colours -- coupled with the conversion table of \cite{pecaut_intrinsic_2013} -- indicative of spectral types M3-M4.5 and absolute brightnesses consistent with the main-sequence. Additionally, the variations were asymmetric, exhibiting a rapid rise and a more gradual decay. This leads us to conclude that the rises we saw were nothing more than standard M dwarf flares in foreground stars. Thus, as expected, we did not detect any microlensing events, despite demonstrating sensitivity to short-duration transient brightenings.


\section{Conclusion}
\label{sec:conclusion}

We have carried out a 67-night photometric monitoring campaign of M3 with the newly commissioned TWIST observatory. The survey has yielded both technical validation of the system and astrophysical results that build on more than a century of variability studies in this cluster. Among the new findings are four SX Phoenicis stars, including one fundamental-mode pulsator that provides an independent distance consistent with published values. The detection of these short-period SXP variables particularly highlights the advantage conferred by TWIST’s high cadence capability and the large data volume of this survey, despite the relatively small 50\,cm aperture. Earlier searches (e.g. \citealt{kaluzny_search_1998,corwin_bv_2001,hartman_bvi_2005,benko_multicolour_2006}), which operated at cadences of order 5--20 minutes and with far fewer data points, were inherently less sensitive to the $\sim$1--3-hour pulsations characteristic of this class.

For the extensive RR Lyrae population, we measured updated periods and amplitudes for the majority of stars, and identified Blazhko-like modulation in over fifty cases. The long timescale of this variability motivated the need for stable photometry over timescales of hundreds of days, which was enabled by the development of a new correction method for scattered-light contamination in flat-field frames.

Notably, amplitudes were established for sixteen RR Lyrae stars that had no published values, improving the completeness of the literature catalogue. We also obtained the first period and amplitude measurements for the unclassified variables V286 and V287, and identified several flaring foreground M dwarfs. A search for microlensing events supported the prediction that the expected rate in M3 is negligible, providing an empirical benchmark for the detectability of these rare events with small-aperture instrumentation. 

Although the analysis presented here uses white-light imaging, TWIST is equipped with a complement of \textit{BVR} filters. Filtered observations were not employed for this programme because the associated reduction in photon flux would have limited our sensitivity to the faintest and lowest-amplitude variables detected in this study. TWIST operates as a general-purpose facility, and since these observations it has been used for a range of projects, including exoplanet transit monitoring, studies of white dwarf variability, and testing of new sCMOS detectors. The circumstances under which the filters will be used in future will depend on the requirements of those programmes, offering flexibility for a wide range of forthcoming projects.

\section*{Acknowledgements}

We would like to thank the anonymous referee for providing valuable feedback which improved this manuscript. This work makes use of data from the TWIST instrument operated on the island of La Palma by the University of Warwick at the Observatorio del Roque de los Muchachos of the Instituto de Astrofísica de Canarias. This work has made use of data from the European Space Agency (ESA) mission
{\it Gaia} (\url{https://www.cosmos.esa.int/gaia}), processed by the {\it Gaia}
Data Processing and Analysis Consortium (DPAC,
\url{https://www.cosmos.esa.int/web/gaia/dpac/consortium}). Funding for the DPAC has been provided by national institutions, in particular the institutions participating in the \textit{Gaia} Multilateral Agreement. This research has made use of the VizieR catalogue access tool, CDS, Strasbourg, France. This research has made use of the SIMBAD database, CDS, Strasbourg Astronomical Observatory, France. MAM and ABC gratefully acknowledge the support of the Science and Technology Facilities Council (STFC) through the training grant ST/X508871/1, which made this research possible. DS is supported by the UK Science and Technology Facilities Council (STFC, grant numbers ST/T007184/1, ST/T003103/1, and ST/T000406/1). IA acknowledges the support of STFC under the CASE Industry scheme ST/W005077/1. 


\section*{Data Availability}
All \textsc{python} scripts, plotting code, extracted light curves, and raw images can be made available upon reasonable request to the author.


\bibliographystyle{mnras}
\bibliography{references}



\appendix

\section{Further characterization}
\label{app:characterization}

\subsection{Dark Current}

\begin{figure}
    \includegraphics[width=8.5cm]{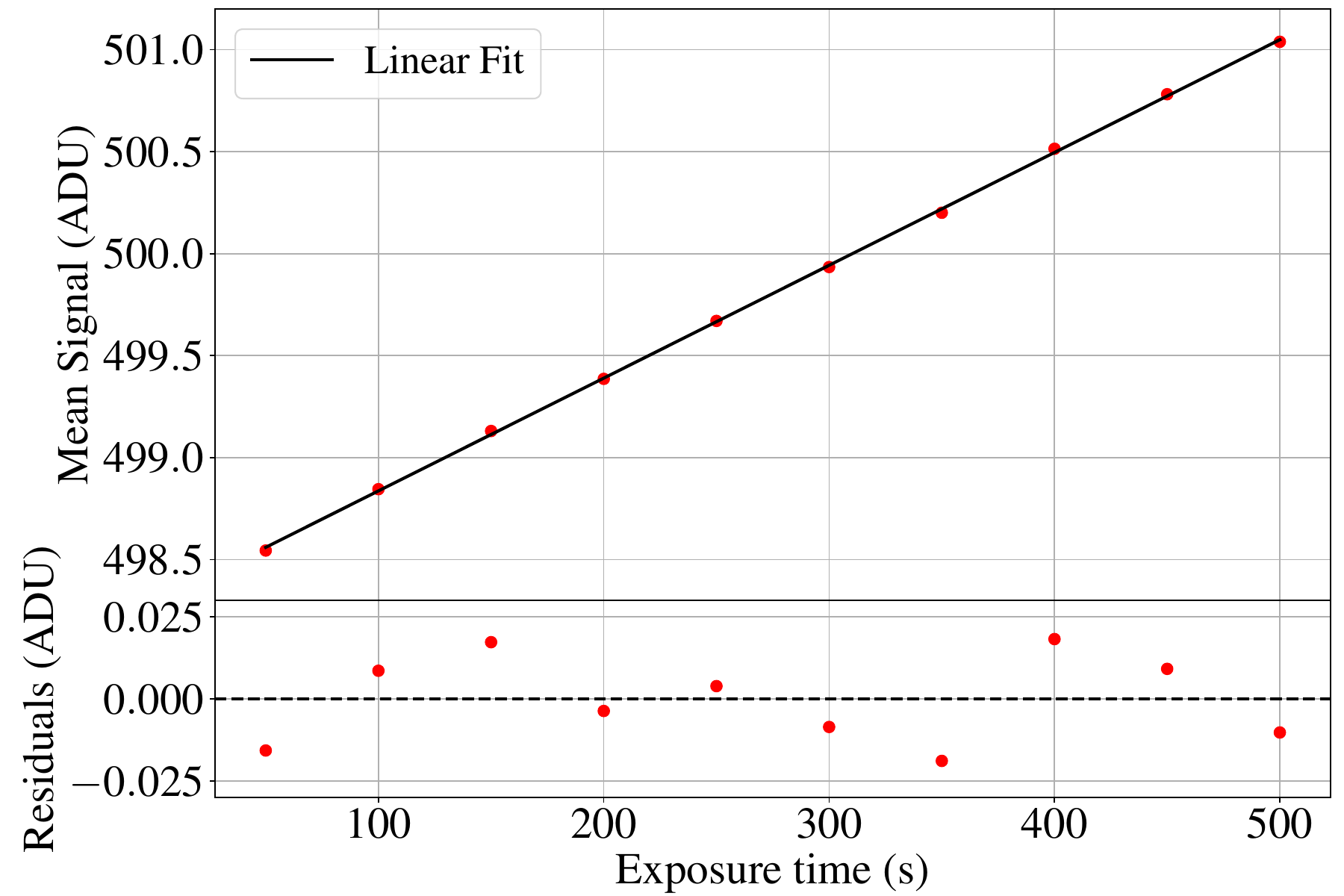}
    \caption{\textbf{Top:} Truncated mean signal versus exposure time for ten dark exposures ranging from 50 to 500 s shown as red points. The mean signal is computed from the central 99 per cent of pixel values, excluding the lowest and highest 0.5 per cent. The black line represents the least-squares linear fit to the data. The gradient of this line is the calculated dark current, and the vertical axis intercept is the mean bias level. \textbf{Bottom:} The residual signal calculated by subtracting the linear model from the data.}
    \label{fig:dark_current}
\end{figure}

Unlike typical CCD cameras, the QHY600M Pro CMOS camera used by TWIST achieves a very low dark current with only modest cooling. We measured the dark current by taking ten dark frames with exposures ranging from 50 to 500\,s under dark conditions with a blocking filter in place and the camera set to 0$^\circ$C. To avoid the influence of dead or hot pixels, we analysed the mean pixel values from the non-overscan region of the images, excluding the smallest and largest 0.5 per cent of values. Fitting a linear model to the mean values, as shown in Fig. \ref{fig:dark_current}, and assuming a gain of 0.414 e$^-$\,ADU$^{-1}$ (see Table \ref{tab:specs}), we calculated a dark current of (8.25\,$\pm$\,0.05) e$^-$\,pixel$^{-1}$\,hour$^{-1}$. This dark current represents a negligible contribution to the overall noise budget, particularly during short (10-second) exposures, for which the expected dark current noise is just 0.151\,e$^-$.

\subsection{Read noise}

\begin{figure}
    \includegraphics[width=8.5cm]{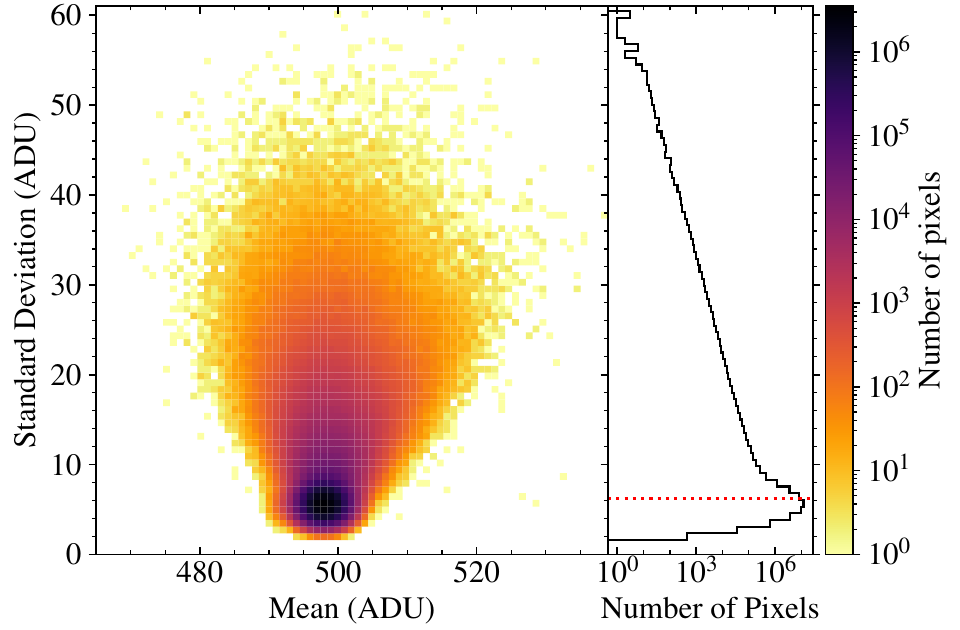}
    \caption{Read noise and mean pixel value distribution from 21 consecutive bias frames. \textbf{Left:} A 2D histogram showing the temporal, frame-to-frame noise and the mean signal of each pixel in these bias frames. The colour bar indicates the number of pixels in each bin. \textbf{Right:} The black histogram shows the distribution of read noise values, and the red dotted line shows the root-mean-square read noise.}
    \label{fig:read_noise}
\end{figure}

With the same setup as above, we took 21 bias images with the shortest available exposure time (39\,$\unit{\us}$) to measure the frame-to-frame variability of individual pixels in the absence of signal -- the read noise. Fig. \ref{fig:read_noise} shows the distribution of the means and standard deviations of the pixel values.

We observe a clearly non-gaussian distribution of read noise values, which can likely be attributed to random telegraph noise (also known as salt and pepper noise), which is characteristic of CMOS sensors and was found to be present in the QHY600M Pro by \cite{alarcon_scientific_2023}. To account for the fact that photometric apertures contain many pixels, we take the overall read noise value to be the root mean of the variances: 2.579 e$^-$. Assuming the dark signal is uniform and follows a Poisson distribution, the read noise is dominant over the dark current noise for exposures shorter than around 48\,minutes.

\subsection{Photon Transfer Curve and Linearity}

The Photon Transfer Curve (PTC) was introduced in \citet{1985SPIE..570....7J} and can be used to extract valuable information about the camera sensor, including the gain, full-well capacity, read noise, and fixed pattern noise. The PTC has two main implementations -- one made with single exposures per signal level and one made with the difference image of two exposures per signal level. The latter approach is more tolerant of less perfectly uniform sensor illumination, although it removes -- and therefore cannot be used to measure -- fixed pattern noise. We adopted this approach based on the testing equipment available to us.

\begin{figure}
	\includegraphics[width=8.5cm]{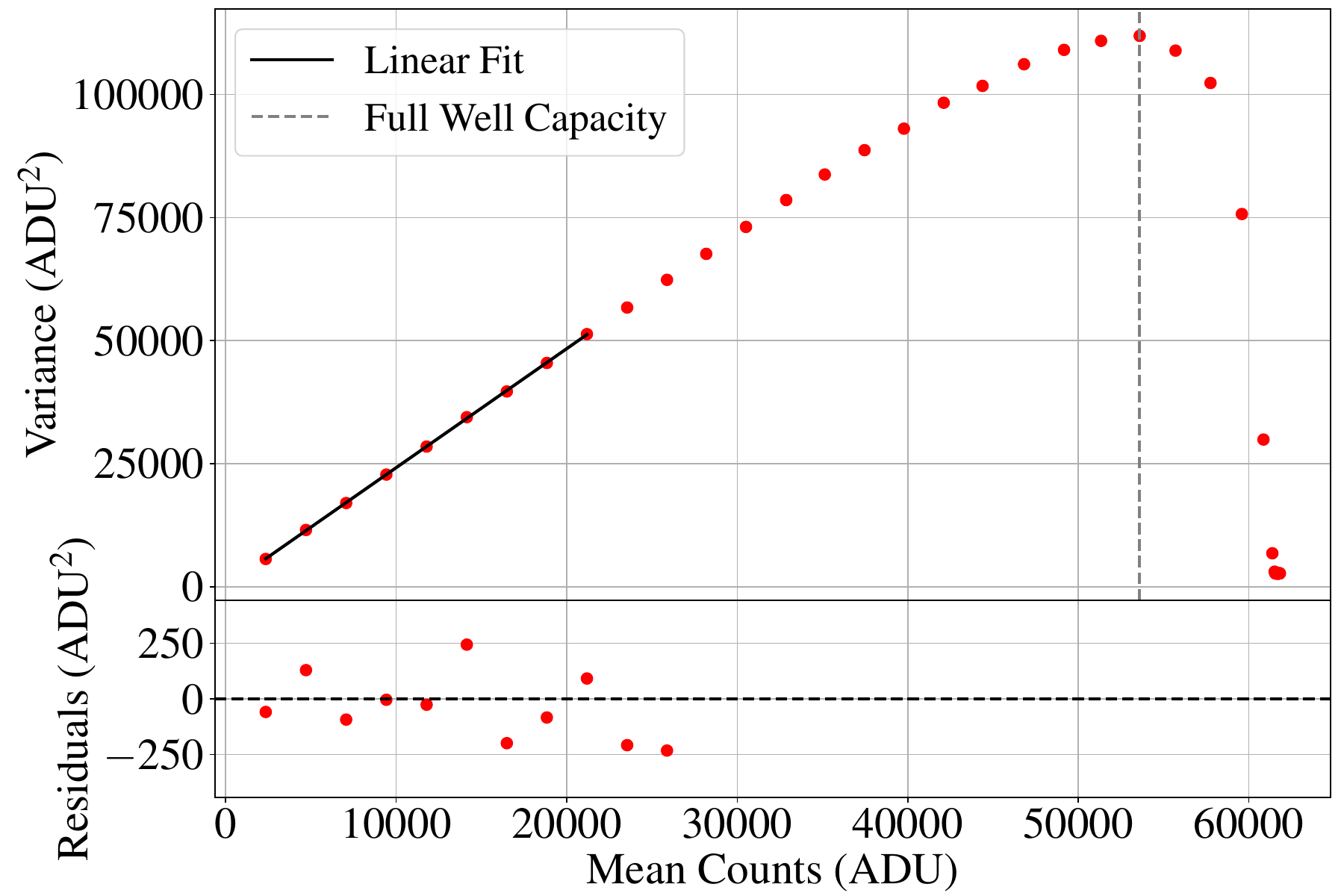}
    \caption{Photon transfer curve for the QHY600M Pro camera in its "Photographic" mode at gain setting 26. The best-fit line in the top plot spans only the data used in the fit. The residuals span the entire data range but quickly fall below the axis limits.}
    \label{fig:ptc}
\end{figure}

We illuminated the sensor with a constant light source in a dark room with the camera off the telescope. We varied the exposure time to achieve different signal levels, taking two images per exposure time. After correcting for bias and dark current, we cropped each image to a 200\,$\times$\,200 pixel central region and measured the mean signal of each pair of images. We then subtracted one image from the other and measured the variance of the difference image. For the PTC, we divide this variance by two to account for the noise contribution from both images in the pair and infer the single-image noise. The PTC is shown in Fig. \ref{fig:ptc}. We fit a straight line to the linear region of the curve, and its inverse gradient yields a gain of 0.414\,e$^-$ADU$^{-1}$ -- very close to the manufacturer's quoted 0.412\,e$^-$ADU$^{-1}$. The full-well capacity is given by the signal level at which maximum variance occurs, around 22,200 electrons. This is much lower than the value of 27,045 electrons given by QHYCCD. In fact, the full-well capacity given by QHYCCD is not possible using the gain they quote, as 16-bit digital saturation would occur first.

\begin{figure}
	\includegraphics[width=8.5cm]{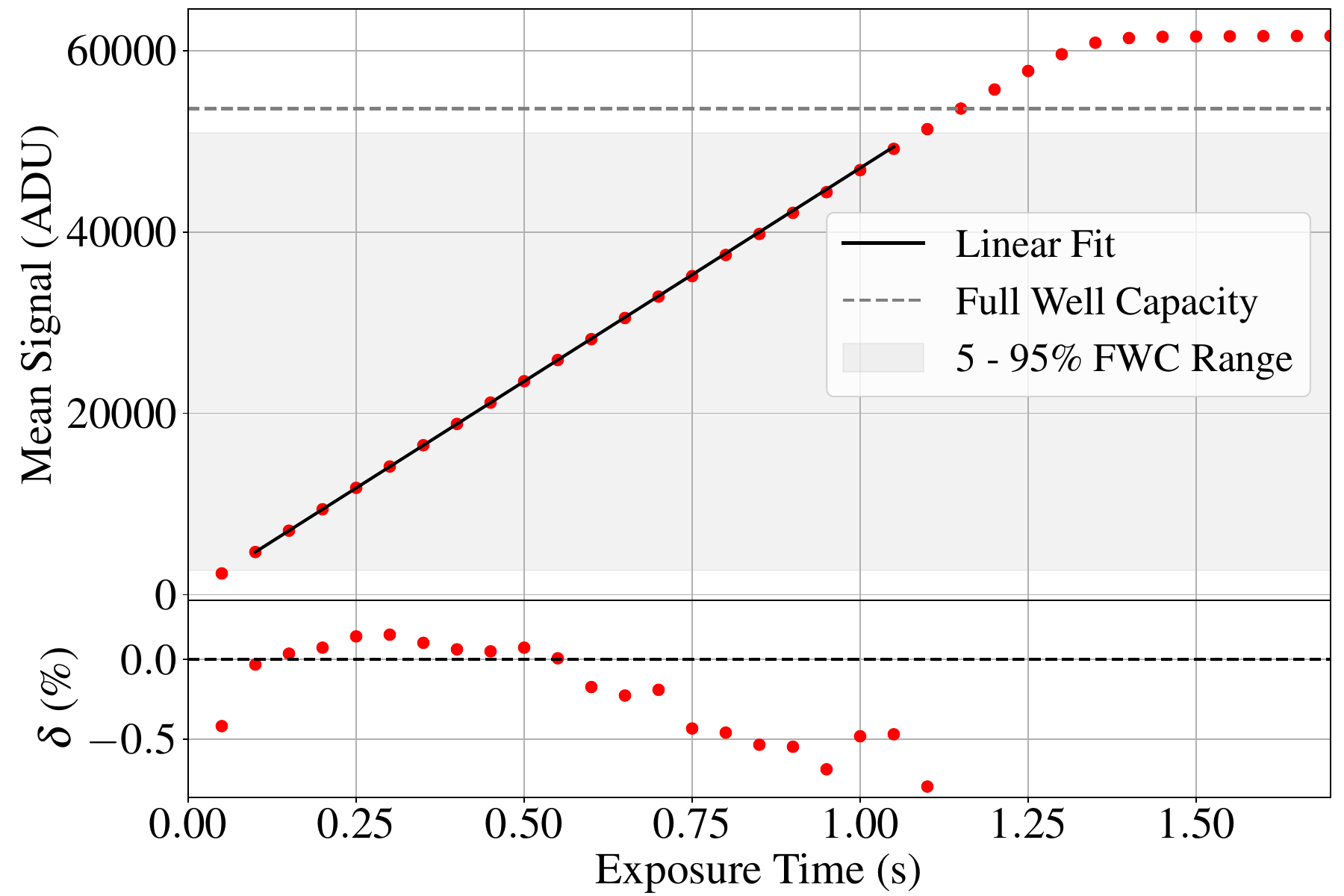}
    \caption{Linearity curve for the QHY600M Pro camera in its "Photographic" mode at gain setting 26. The best-fit line in the top plot, which minimizes the absolute sum of the \textit{relative} residuals, spans only the data used in the fit. The residuals span the entire data range but quickly fall below the axis limits. $\delta$ expresses the relative residuals as a percentage.}
    \label{fig:linearity}
\end{figure}

With the same set of measurements used for the PTC, but instead using only one image per exposure level, we measured the linearity error of the sensor according to version 4.0 of the EMVA 1288 standard \citep{jahne2010emva}. Therein, the linearity error is calculated by considering a linear regression of, with respect to time, the signal within a range of 5-95 per cent of the full-well capacity. The regression minimizes the \textit{relative} residuals -- the residuals divided by their respective best-fit values. The linearity error is then defined as the mean of the absolute values of these relative errors, in this case 0.248 per cent. The linearity curve is shown in Fig. \ref{fig:linearity}.

\subsection{Spectral Response}

\begin{figure}
	\includegraphics[width=8.5cm]{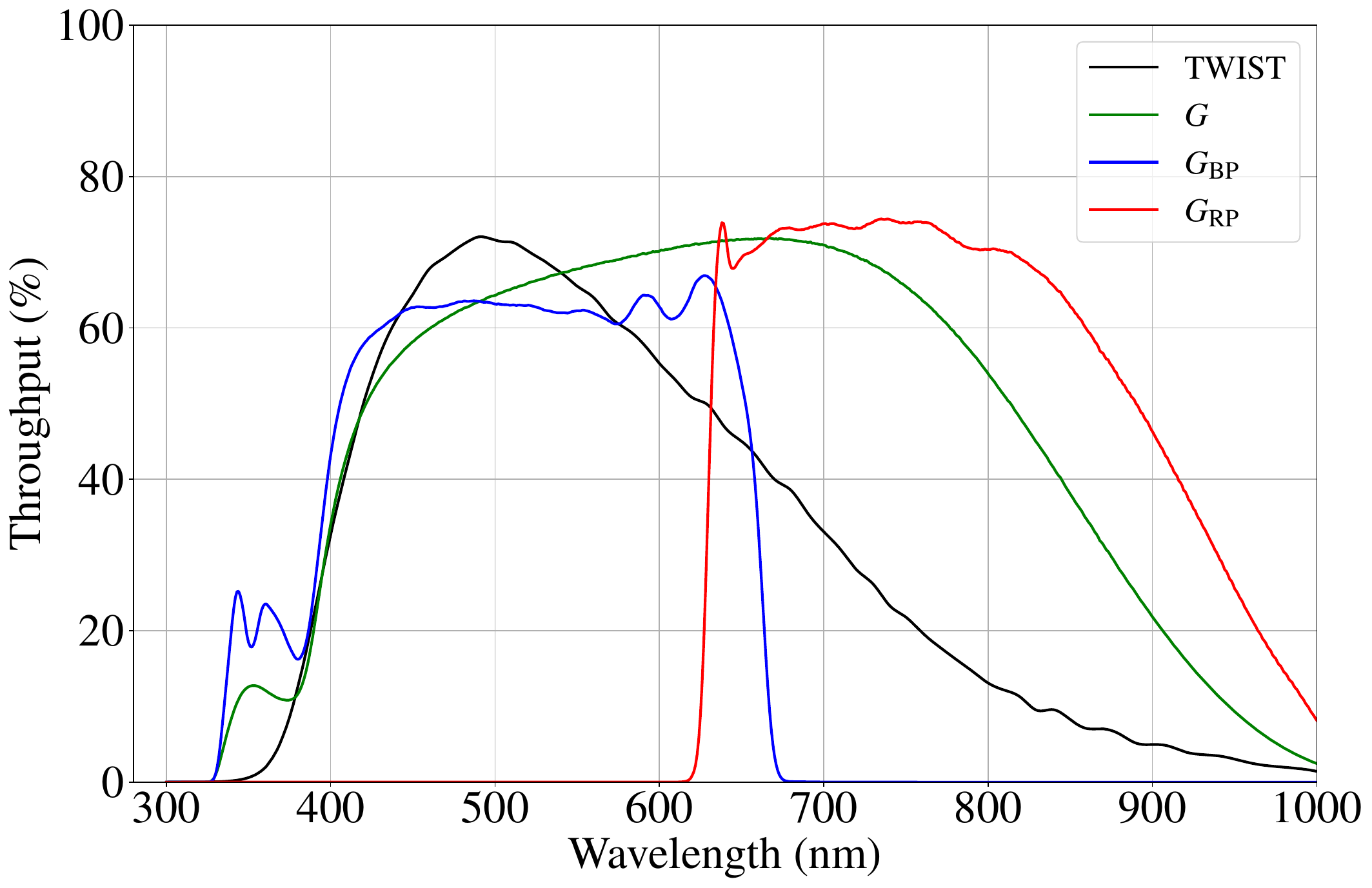}
    \caption{Optical throughput of TWIST and \textit{Gaia}. The black line represents the modelled total throughput of TWIST, including quantum efficiency and camera lens transmission taken from \protect\cite{gill_low-cost_2022}, atmospheric extinction at the Roque de Los Muchachos observatory at an airmass of 1 from \protect\cite{king_atmospheric_1985}, and assuming a constant transmission of 93 and 99.5 per cent for the mirrors and corrector lenses respectively. The green, blue and red lines show respectively the \textit{Gaia} (E)DR3 \textit{G}, $\textit{G}_\text{BP}$, and $\textit{G}_\text{RP}$ passbands from \protect\cite{riello_gaia_2021}.}
    \label{fig:transmission}
\end{figure}

Like most modern sCMOS detectors, the Sony IMX455 sensor of the QHY600M Pro is most sensitive to blue light, reaching a peak quantum efficiency of over 90 per cent between 450 and 510\,nm according to QHYCCD. A model of the optical throughput of the entire system, compared to the \textit{Gaia} DR3 passbands \citep{gaiadr3}, is shown in Fig. \ref{fig:transmission}. PlaneWave Instruments does not publicly share the wavelength-dependent transmission of the lenses and mirrors of the OTA. Hence, we approximate the throughputs of each mirror and each lens to be constant at 93 and 99.5 per cent, their averages over the interval 400--700\,nm. We note that the transmission is likely lower outside that range.

\begin{figure}
        \includegraphics[width=8.5cm]{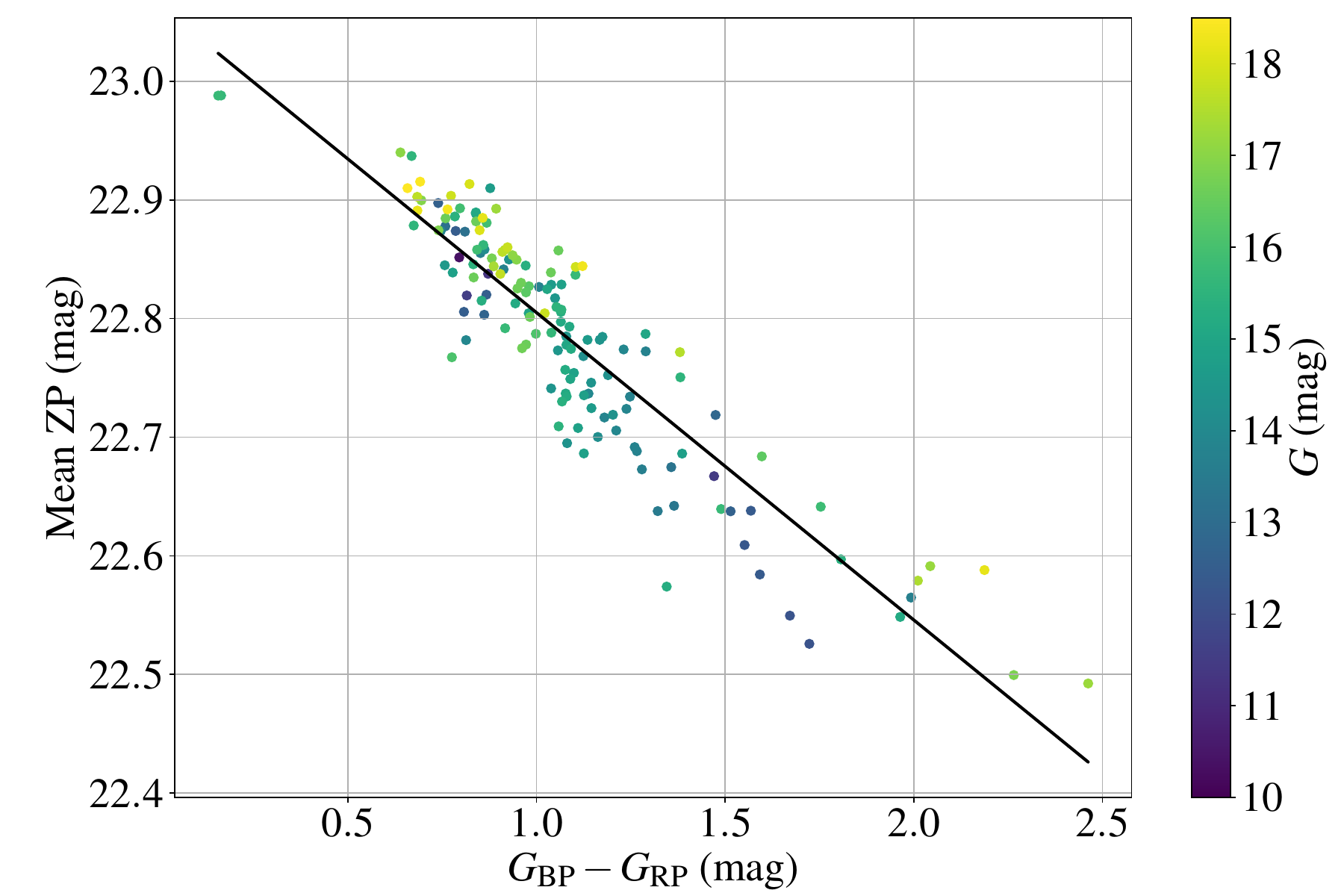}
    \caption{Mean \textit{G}-band zeropoint magnitudes (difference between \textit{G} magnitudes and flux-averaged instrumental magnitudes for a 5-pixel aperture radius) versus $\textit{G}_\text{BP} - \textit{G}_\text{RP}$ colours for 148 stars brighter than \textit{G}\,=\,18.5. The stars chosen are not marked as variables in \textit{Gaia} DR3. They are unblended, which we define as being at least 22.5 arcsec (50 binned pixels) from a source brighter than three magnitudes dimmer than that star. The data points are coloured according to the stars' catalogued \textit{G} magnitudes. A 3$\sigma$-clipped linear least squares fit to the data is overplotted.}
    \label{fig:flux-vs-mag}
\end{figure}

Additionally, we examined the spectral and absolute response of the system observationally using the unblended, non-variable stars in the survey. Fig. \ref{fig:flux-vs-mag} shows the measured flux-averaged \textit{G}-band zeropoint magnitudes of the stars against their $\textit{G}_\text{BP}-\textit{G}_\text{RP}$ colours. The trend is approximately linear over the colour range of the sample and shows a clear reduction in sensitivity towards redder stars. Applying a 3$\sigma$-clipped linear regression to the individual zeropoints yields an average colour-dependent \textit{G}-band zeropoint, 

\begin{equation}
\text{ZP}_\textit{G}=23.0640-0.2591\times(\textit{G}_\text{BP}-\textit{G}_\text{RP})\pm0.0413.    
\end{equation}

The scatter about the fit is almost entirely not due to measurement uncertainties, but rather a systematic uncertainty arising from the complexity of the real stellar spectra that cannot be captured by the $\textit{G}_\text{BP}-\textit{G}_\text{RP}$ colour alone. We note that this calculation uses fluxes from 5-pixel radius apertures, which capture, on average, 70.6 per cent as much flux as 15-pixel radius apertures in our testing -- equivalent to a 0.388 mag lower zeropoint.

\section{Photometric precision}
\label{app:photometric_precision}

The photometric performance of the system was measured from observations of the globular cluster M3 and compared to a theoretical noise model. The noise model combines errors from the target, sky background, dark current, read noise, and scintillation. The total noise for a point source, $N_{\text{total}}$, is modeled by equation \ref{eq:noise_model}:

\begin{equation}
    N_{\text{total}}=\sqrt{F_{\text{target}} + n_{\text{pixel}}f_{\text{sky}} + n_{\text{pixel}}f_{\text{dark}} + n_{\text{pixel}}\sigma^2_{\text{read}} + \sigma^2_\text{Y}F_\text{target}^2},
    \label{eq:noise_model}
\end{equation}

where $F_\text{target}$ is the total integrated target flux in electrons, $f_\text{sky}$ and $f_\text{dark}$ are the integrated sky and dark flux in electrons per pixel, $n_\text{pixel}$ is the number of pixels in the photometry aperture, $\sigma_\text{read}$ is the read noise of one pixel in electrons, and $\sigma_\text{Y}$ is the fractional scintillation noise approximated by the modified Young formula \cite[]{young_photometric_1967,osborn_atmospheric_2015}:

\begin{equation}
    \sigma_\text{Y}^2=10^{-5} C^2_\text{Y}D^{-4/3}t^{-1}\sec^{3}{z}\exp{(-2h_\text{obs}/H)},
\end{equation}

where $C^2_\text{Y}$ is the empirical Young's coefficient that depends on the level of atmospheric turbulence in the vicinity of the observatory site. $D$ is the telescope aperture size, $t$ is the exposure integration time, $z$ is the zenith angle ($\sec{z}$ is the approximate airmass), $h_\text{obs}$ is the height of the observatory above sea level, and $H$ is the atmospheric scale height. The relevant values of the quantities described above for the TWIST/La Palma are listed in Table \ref{tab:obs_quantities}. 

\begin{table}
	\centering
	\caption{Relevant quantities influencing the noise properties of the photometric system within our model. The dark current and read noise were measured in the present work using our chosen camera setup of 0$^{\circ}$C and gain setting 26 in Photographic mode.}
	\label{tab:obs_quantities}
	\begin{tabular}{lll}
		\hline
		\textbf{Quantity} & \textbf{Value} & \textbf{Reference}\\
		\hline
		  $H$                  & 8000 m                         & \cite{osborn_atmospheric_2015} \\
            $C_\text{Y}$         & 1.30 m$^{2/3}$s$^{1/2}$        & \cite{osborn_atmospheric_2015} \\
            $f_\text{dark}$      & 0.00229 e$^-$pixel$^{-1}$s$^{-1}$ & This work \\
		  $\sigma_\text{read}$ & 2.579 e$^-$                    & This work\\
            $D$                  & 0.508 m                        & \\
            $t$                  & 10 s                           & \\
            $h_\text{obs}$       & 2349 m                         & \\
		\hline
	\end{tabular}
\end{table}

\begin{figure}
	\includegraphics[width=8.5cm]{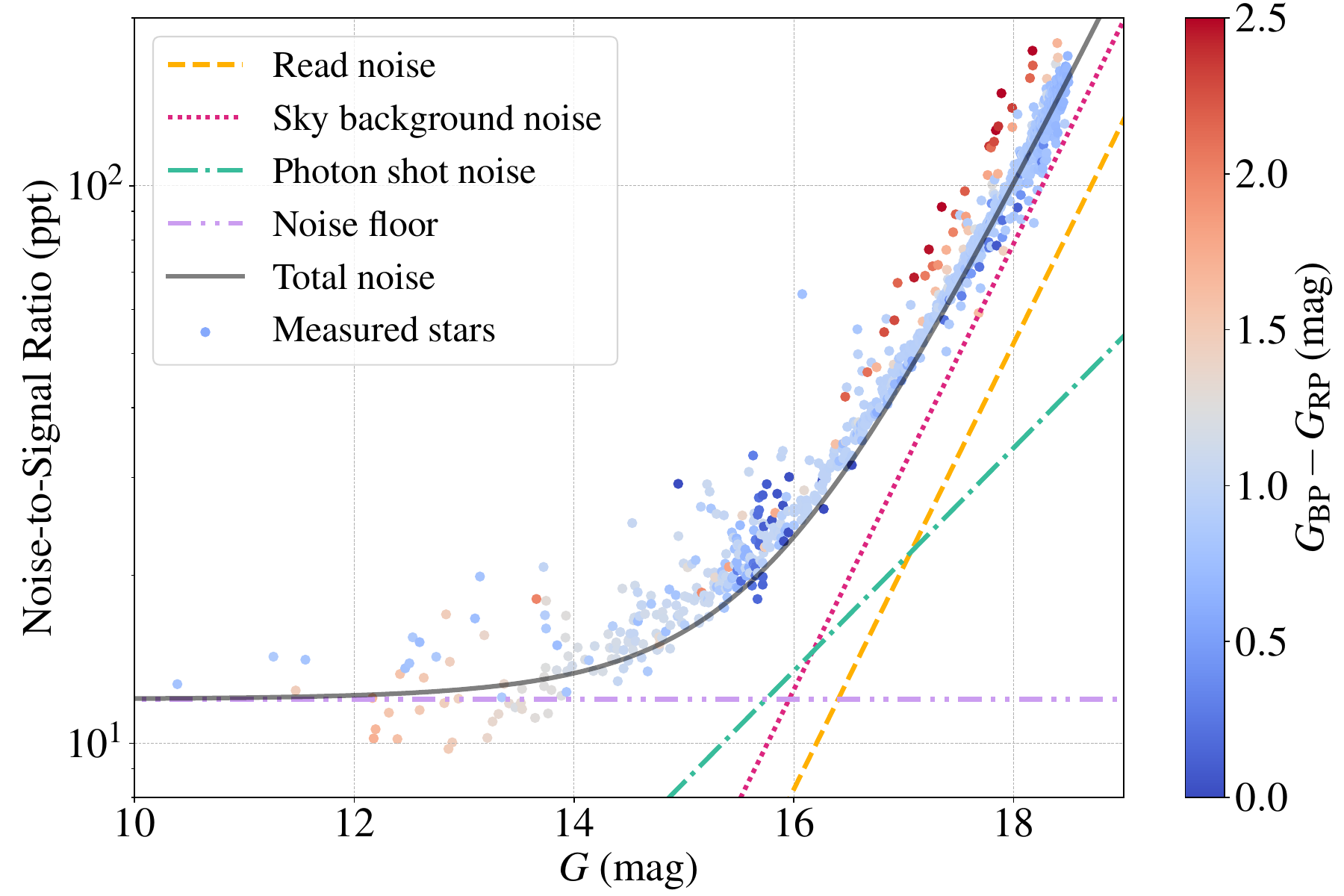}
    \caption{Photometric precision as a function of stellar magnitude predicted by our noise model and measured by our observations at a 10-second cadence. The grey line represents the total modelled noise, with contributions from photon shot noise (green), sky background noise (pink), and read noise (orange). The empirical noise floor component is shown in purple. Dark current noise is negligible and would not be visible on this scale. The dots represent the measured scatter in the detrended light curves of 993 stars categorized in \textit{Gaia} DR3 as non-variable and brighter than \textit{G}\,=\,18.5. The noises are scaled by the inverse square root of the proportion of images in which a given star is detected to portray the precision achievable in a fixed time interval. The points are coloured by the stars' $\textit{G}_\text{BP}-\textit{G}_\text{RP}$ colours.}
    \label{fig:nsr}
\end{figure}

We found that the photometric aperture size chosen to maximize the signal-to-noise ratio for dim stars is not ideal for brighter stars. This is likely for several reasons. Firstly, averaged across all the images used for photometry, 5-pixel radius apertures captured only 70.6 per cent as much flux as 15-pixel radius apertures (the largest we tested). As a result, the "signal" component of the SNR is diminished. This is especially significant for sources whose precision is limited by photon shot noise. Secondly, the use of small photometric apertures exacerbates the impact of their imperfect placement; a difference in the placement of the aperture produces a larger difference in the collected flux for a smaller aperture. Finally, and most pertinent to the brightest stars, small apertures prevent spatial averaging over distortions of the stars' PSFs due to atmospheric scintillation. As a result, the theoretical scintillation noise floor was not reached, except in our testing, where we used much larger 15-pixel radius apertures. Hence, we alter the model, replacing the scintillation component with the $\sim$12 ppt empirical noise floor we observed. The model and its constituent components are shown alongside the measured noise-to-signal values in Fig. \ref{fig:nsr}.

Even with the very small pixel angular scale present in our setup, the per-pixel read noise is low enough that sky background dominates at larger magnitudes. Photon shot noise is only dominant at around $G=16$. As expected, redder stars have greater noise than bluer stars of the same \textit{G} magnitude, owing to TWIST's blue-biased spectral response.

For simplicity, the preceding analysis neglects non-uniform response across the field arising from vignetting. In the current optical configuration, the flat-field response at the dimmest corner of the detector is approximately 80.6\% of that at the brightest point near the image centre. To assess the impact of this variation, we generated two additional noise models -- representing the brightest and dimmest regions of the field -- by modulating the zeropoints and sky background levels of the nominal model according to the relative flat-field response. The resulting predictions indicate that bright stars already at the noise floor experience identical photometric precision across the field. For photon-limited stars, the expected increase in noise from the centre to the edge of the detector is 11.4\%, while for sky-background-limited stars, the increase rises asymptotically with magnitude towards 15.1\%.

As shown in Fig. \ref{fig:obs}, the median PSF half-flux diameter over the survey is around 3 arcsec. The best image achieved 1.7 arcsec. This is significantly higher than the typical atmospheric seeing from La Palma and the theoretical diffraction limit for a 0.5\,m class telescope. It is also worse than the 1.3 arcsec achieved by pt5m \citep{hardy_pt5m_2015}, another 50\,cm telescope located just 215 metres away from TWIST. The larger-than-expected PSF has persisted despite attempts to recollimate the telescope and tune the mount; we suspect dome seeing, perhaps related to the slow release of heat from the concrete floor, to be a likely culprit. The poor seeing has a significant, negative impact on the signal-to-noise ratio of dim stars due to the need for larger photometry apertures, capturing more read noise and sky-background noise. We would expect the SNR to double for stars in this regime if the typical seeing were instead around 1.5 arcsec.


\bsp	
\label{lastpage}
\end{document}